\documentclass[10pt,journal,cspaper,compsoc]{IEEEtran}

\usepackage{amsfonts}
\usepackage{amsmath}
\usepackage{amssymb}
\usepackage{amsthm}
\usepackage{graphicx}
\usepackage{multirow}
\usepackage{listings}
\usepackage{color}
\usepackage{xcolor}
\usepackage{framed}
\ifCLASSOPTIONcompsoc
\else
\fi

\ifCLASSINFOpdf
\else
\fi

\newcommand{\tabincell}[2]{\begin{tabular}{@{}#1@{}}#2\end{tabular}}

\begin{document}

%--
\title{Enabling Context-awareness by Predicate Detection in Asynchronous Pervasive Computing Environments}

\author{Yiling~Yang,
        Yu~Huang,
        Xiaoxing~Ma,
        and~Jian~Lu
    \IEEEcompsocitemizethanks{

        \IEEEcompsocthanksitem Corresponding author: Yu Huang, State Key Laboratory for Novel Software Technology, and Department of Computer Science and Technology, Nanjing University, Nanjing, China, 210046.\protect\\
        E-mail: yuhuang@nju.edu.cn.

        \IEEEcompsocthanksitem Yiling Yang, Xiaoxing Ma, and Jian Lu are with the State Key Laboratory for Novel Software Technology, and Department of Computer Science and Technology, Nanjing University, Nanjing, China, 210046.\protect\\
        E-mail: csylyang@gmail.com, \{xxm, lj\}@nju.edu.cn.
    }
}

% The paper headers
%\markboth{IEEE TRANSACTIONS ON SOFTWARE ENGINEERING,~Vol.~X, No.~X, XXX~XXXX}%
%{Shell \MakeLowercase{\textit{et al.}}: Enabling Context-awareness by Predicate Detection in Asynchronous Pervasive Computing Environments}

\IEEEcompsoctitleabstractindextext{
\begin{abstract}
  Pervasive applications are involving more and more autonomous computing and communicating devices, augmented with the abilities of sensing and controlling the logical / physical environment. To enable context-awareness for such applications, we are challenged by the intrinsic asynchrony among the context collecting devices. To this end, we introduce the predicate detection theory and propose the Predicate-Detection-based Context-Awareness (PD-CA) framework, in which: a) logical time is used to explicitly cope with the asynchrony; b) specification of predicates enables the applications to express contextual properties of their concerns; c) online and incremental predicate detection algorithms effectively enable context-awareness at runtime. Under the guidance of the PD-CA framework, we present the design and implementation of the MIPA middleware, which shields the applications from the burden of processing the asynchronous contexts. We also demonstrate how PD-CA simplifies the development of context-aware applications. Experimental evaluations show the performance of MIPA in supporting context-aware applications despite of the asynchrony.
\end{abstract}

%--
}

%--
\maketitle

%--
\IEEEdisplaynotcompsoctitleabstractindextext
\IEEEpeerreviewmaketitle

%========================================================================
\section{Introduction}\label{sec:Introduction}

Pervasive applications are undergoing changes as more and more mobile devices are augmented with sensing and controlling abilities, besides the basic abilities of computation and communication. We call such devices C$^3$S (\underline{C}omputation, \underline{C}ommunication, \underline{C}ontrol, and \underline{S}ensing) devices. Examples of C$^3$S devices include mobile robots patrolling in a chemical plant for safety management \cite{Zhan13, Duggirala12} and smart phones equipped with a variety of sensors \cite{Lane10, Azizyan09}.

C$^3$S devices can provide rich context information for the applications, and pervasive applications are typically designed to be context-aware, i.e., intelligently adapting their behavior to the environment \cite{Dey00, Lu08, Yang08}. However, enabling context-awareness through C$^3$S devices is faced with severe challenges, as detailed below.

The contexts of interest to a pervasive application often span a geographically large area, and contain rich semantics. This is often beyond the ability of one single C$^3$S device. Thus, a group of autonomous but also coordinating C$^3$S devices should be deployed. Take a chemical plant scenario for example. A group of mobile robots are deployed to periodically patrol the plant for safety management \cite{Zhan13, Duggirala12}. The robots need to proceed in certain formation to cover all possible spots of hazardous material leak. Appropriate spreading of multiple robots can also enable the robots to collect contexts with better quality, e.g., to sense the average temperature in the plant.

The coordination among the C$^3$S devices is intrinsically asynchronous. There is no global clock available among the C$^3$S devices. Constrained resources and task scheduling of the C$^3$S devices (often embedded systems) may lead to unpredictable computation delay \cite{Kshemkalyani13}. The growing adoption of wireless communications, which are prone to bandwidth shortage, network congestion, unpredictable routings, and retransmission, leads to unpredictable communication delay \cite{Schneider93, Patt94, Gartner99, Huang12}. All these characteristics of the C$^3$S devices and their communication networks lead to the intrinsic asynchrony among the contexts they collect \cite{Huang12, Kshemkalyani13}.

One possible solution to cope with the asynchrony is clock synchronization. However, clock synchronization may not enable correct and fault-tolerant coordination among the autonomous C$^3$S devices \cite{Gartner99, Schneider90, Duggirala12}. Thus, it cannot enable context-awareness despite of the asynchrony in pervasive scenarios enriched of coordinating C$^3$S devices. Specifically, each C$^3$S device only has its own local clock, which cannot be perfectly synchronized \cite{Patt94, Corbett12, Li06, Duggirala12}. The uncertainty caused by the skew among the clocks may lead to incorrect behavior \cite{Kshemkalyani13}. Besides, clock synchronization schemes make assumptions on process execution speeds and communication delay \cite{Patt94}. These assumptions may not be guaranteed for the autonomous C$^3$S devices. A group of robots are prone to incorrect behavior even if one single assumption is violated \cite{Gartner99}. The inaccuracy of synchronization and the potential violation of assumptions make reasoning based on time and timeouts a delicate and error-prone undertaking \cite{Gartner99}. Furthermore, periodic clock synchronization may be unaffordable in terms of energy consumption, or be hampered due to device autonomy and administrative boundaries such as privacy concerns and security issues \cite{Kshemkalyani13, Huang12}. Consequently, it is more practical to have few or, better, no synchrony assumption in scenarios of coordinating C$^3$S devices \cite{Duggirala12, Kshemkalyani13, Gartner99}.

To enable context-awareness for pervasive applications enriched with C$^3$S devices, we introduce the predicate detection theory and propose the Predicate-Detection-based Context-Awareness (PD-CA) framework, which consists of three essential parts:
%--
\begin{itemize}
  \item Logical time is used to cope with the asynchrony among contexts collected from the system of autonomous C$^3$S devices. Temporal orders among asynchronous contextual events are encoded and decoded via the logical vector clock. Global snapshot of the asynchronous system of C$^3$S devices is redefined under the notion of logical time. Dynamic behavior of the C$^3$S devices are modeled over sequences of global snapshots.
  \item Specification of predicates enables the applications to express their concerns on properties of the contexts. Based on the modeling above, the specification can delineate local contextual properties on local states of one C$^3$S device, global contextual properties on snapshots of the system of C$^3$S devices, and dynamic behavioral properties on sequences of snapshots.
  \item Context-awareness is enabled by detection of the specified contextual property at runtime. The detection is such a persistent process that new emerging contexts trigger the incremental detection of the specified property.
\end{itemize}

Under the guidance of the PD-CA framework, we develop the Middleware Infrastructure for Predicate detection in Asynchronous environments (MIPA) \cite{MIPA}. In context-aware computing scenarios, MIPA first receives contextual properties from the applications. It then decomposes the global contextual properties to local ones, with which MIPA instructs each C$^3$S device to collect the related contexts. MIPA detects the specified contextual properties with the contexts in an online and incremental manner and informs the applications when the properties are satisfied. MIPA adopts a layered architecture to support this context processing process.

We further discuss how MIPA shields the applications from the burden of coping with asynchronous contexts. Based on the PD-CA framework, the context-aware adaptation logic of the application is constructed in a condition-action manner. The contextual property serves as the condition of context-aware behavior.
%Collection of asynchronous contexts and detection of the contextual properties are conducted by MIPA. When MIPA detects a contextual property true, the corresponding context-aware behavior of the application will be triggered.
Experimental evaluations show the performance of MIPA in supporting context-aware applications despite of the asynchrony.

The rest of this work is organized as follows. Section \ref{sec:Motivation} presents a motivating example. Section \ref{sec:PD theory} overviews preliminaries of the predicate detection theory. Section \ref{sec:PD approach} presents our PD-CA framework. Section \ref{sec:Architecture and implementation} discusses the design of MIPA. Section \ref{sec:develop app} discusses how MIPA simplifies the development of context-aware applications. Section \ref{sec:Experiments} presents the experimental evaluation. Section \ref{sec:Related work} discusses the related work. In Section \ref{sec:Conclusion}, we conclude the work and discuss the future work.

%========================================================================
\section{Motivations and Challenges by Example}\label{sec:Motivation and challenges}

To further justify our PD-CA framework, we first introduce a case study scenario. Then we discuss the motivations and challenges in detail based on this scenario.

%------------------------------------------------------------------------
\subsection{Chemical Plant Safety Management Scenario}\label{sec:Case study}

\begin{figure}[tbp]
  \centering
  \includegraphics[width=3.2in]{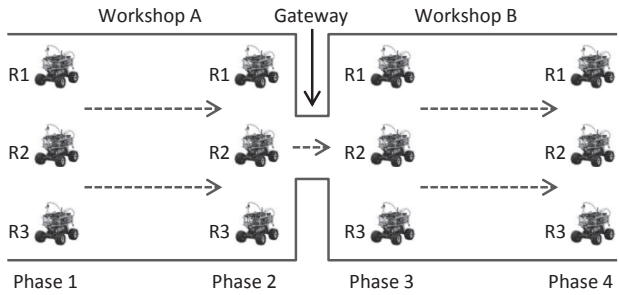}
  \caption{Robots patrolling in a chemical plant.}
  \label{F:Patrolling a factory}
\end{figure}

Let us consider a safety management application of a chemical plant, as shown in Fig. \ref{F:Patrolling a factory}. Several mobile robots are deployed to periodically patrol the plant. Robots are equipped with temperature sensors and hazardous material leak sensors for safety management. To ensure the safety of the plant, the application is concerned with the following contextual properties:
%--
\begin{itemize}
  \item $\phi_1$: all the robots detect hazardous material leak in a workshop.
  \item $\phi_2$: the average temperature in a workshop gets exceptionally high.
\end{itemize}

\noindent When $\phi_1$ is detected, the safety management application should sound the alarm, notify the workers in the workshop to escape, and notify the plant safety officer nearby to cope with the leak. When $\phi_2$ is detected, the safety management application should notify the fireman nearby to cope with the possible fire accident.

To ensure the quality of surveillance, the application is also concerned with the status of the coordination among the robots. For example, the application is concerned with whether the robots proceed in the right formation (to evenly cover a workshop). It is also concerned with whether the robots pass the gateway one by one in certain order (since only one robot can pass the gateway at the same time, the robots need to coordinate to pass the gateway one by one). Specifically, the application is concerned with the following properties:
%--
\begin{itemize}
  \item $\phi_3$: robot R1 or R3 cannot detect R2 and R2 cannot detect R1 or R3.
  \item $\phi_4$: the robots pass the gateway one by one, in the order of R1, R2, and R3.
\end{itemize}

\noindent When $\phi_3$ is detected, the safety management application should notify the robots to restore the formation before proceeding. When $\phi_4$ is violated, the safety management application should notify the robots to prevent congestion at the gateway.
%--

%------------------------------------------------------------------------
\subsection{Motivations}\label{sec:Motivation}

In this safety management scenario, local clocks of the robots cannot be perfectly synchronized. Limited computation capacities and reliance on wireless communications lead to computation and communication delay \cite{Kshemkalyani13, Huang12}. The asynchrony among these mobile robots makes it a delicate and error-prone undertaking to reason both the contexts collected from the robots and the coordination status of the robots based on time and timeouts.

For example, the robots coordinate to pass the gateway one by one in certain order. They could first synchronize their clocks. Then the order among the robots can be easily achieved based on synchronized time. However, the synchronization relies on multiple assumptions. If even one single assumption is violated, the robots may behave incorrectly, e.g., contend and collide around the gateway.

Moreover, in such dynamic and uncertain environments, we may deploy redundant robots to achieve fault-tolerance, e.g., using the replicated state machine approach \cite{Schneider90, Gartner99}. Thus, the robots and their backups first need to achieve agreement, i.e., guaranteeing that every replica sees the same set of events. Then the robots and their backups need to achieve order, i.e., guaranteeing that every replica sees the events in the same order \cite{Schneider90}. However, clock synchronization has intrinsic inaccuracy and may lead to errors in maintaining the order \cite{Schneider90}.

%------------------------------------------------------------------------
\subsection{Challenges}\label{sec:Challenges}

In order to provide precise and reliable contexts based on C$^3$S devices, it is more practical to have few or no synchrony assumptions in such scenarios. Consequently, the mobile robots in this scenario should coordinate based on the asynchronous message-passing model \cite{Lamport78}. Application developers can develop asynchronous algorithms for the robots to coordinate and to fulfil their tasks. In order to achieve context-awareness through a group of asynchronously coordinating mobile robots, there are mainly three critical challenges.

{\it Challenge 1. How to interpret the asynchronous contexts?} The context model needs to explicitly cope with the asynchrony among contexts. Specifically, the application is often concerned with global properties of the environment, while the related contexts are generated from the distributed mobile robots. Because the robots coordinate in an asynchronous manner, the contexts generated from them are also asynchronous. Take property $\phi_3$ in Section \ref{sec:Case study} as an example. Without global clocks, we cannot tell whether the three pieces of contexts ``R1 cannot detect R2'', ``R3 cannot detect R2'', and ``R2 cannot detect R1 or R2'' occur concurrently. Existing context modeling techniques mainly assume the availability of global clocks, but cannot cope with distributed and asynchronous contexts \cite{Xu05, Xu10, Ranganathan03a}. Thus, the precise modeling of distributed and asynchronous contexts is required, which enables further specification and detection of contextual properties the application is concerned with.

{\it Challenge 2. How to enable the application to express its concerns on the asynchronous computing environment?} A specification formalism is required, which can enable the application to delineate its concerns on the asynchronous contexts, and inform the related robots of their tasks of context acquisition. Specifically, the application may specify various types of contextual properties. The properties may describe status of the physical environment, such as hazardous material leak and indoor temperature. The properties may also describe status of the asynchronous coordination among the robots, such as whether the robots correctly pass the gateway. Based on the specification, we can accurately and conveniently tell the robots what contexts they should collect. The application is then shielded from the underlying details of acquisition and processing of asynchronous contexts.

{\it Challenge 3. How to efficiently detect changes of the computing environment?} Online and incremental detection algorithms are required to detect the contextual properties. As shown in the scenario, the application has close interaction with the robots at runtime. The detection algorithms are thus required to be online. The detection should also be finished as soon as possible, when the property of interest to the application occurs. Thus, the detection algorithms should be incremental, only checking the newly collected contexts. Processing all the contexts each time is cost-ineffective and often unnecessary, and should be avoided.

%========================================================================
\section{Preliminaries of Predicate Detection}\label{sec:PD theory}

The PD-CA framework relies on the theory of predicate detection \cite{Mattern89, Cooper91, Schwarz94, Babaoglu93, Garg94, Garg96, Babaoglu96, Huang12, Yang13, Sen07, Yang13a}. Predicate detection refers to the checking of global predicates over asynchronous computations \cite{Cooper91}. Predicate detection consists of three essential parts: 1) modeling of the asynchronous computation; 2) specification of global predicates; 3) detection of specified predicates \cite{Yang13, Huang12}. We outline these three parts below.

%------------------------------------------------------------------------
\subsection{Modeling of the Asynchronous Computation}\label{sec:Modeling}

A distributed system consists of a collection of {\it instrumented processes} $P^{(1)}$, $P^{(2)}$, $\cdots$, $P^{(n)}$. Examples of instrumented processes include a thread in a program or a software process manipulating a C$^3$S device. One {\it checker process} $P_{che}$ is in charge of collecting the trace of the distributed system execution and detecting the specified predicate.

%------------------------------------
\subsubsection{Logical Time}\label{sec:asynchronous message-passing}

We model the processes as a loosely-coupled message-passing system, without any global clock or shared memory. Communications suffer from finite but arbitrary delay. Different processes may run at different speeds. We assume that dedicated message protocols are employed to ensure that no messages are lost, altered, or spuriously introduced \cite{Garg94, Garg96}.

As the system executes, each $P^{(k)}$ generates its (potentially infinite) trace of {\it local states} connected by {\it events}: ``$s^{(k)}_0, e^{(k)}_1, s^{(k)}_1, e^{(k)}_2, \cdots$''. The order among the events of instrumented processes is encoded by the logical vector clock \cite{Mattern89} which indicates the {\it happen-before} relation (denoted by `$\rightarrow$') resulting from message causality \cite{Lamport78}. Detailed definitions of the happen-before relation and the logical vector clock can be found in \cite{Schwarz94, Yang13, Huang12, Garg96, Babaoglu96}. For example, we have $s^{(1)}_{0} \rightarrow s^{(2)}_{2}$ and $s^{(2)}_{0} \rightarrow s^{(1)}_{2}$ in Fig. \ref{F:STD and lattice}.

%------------------------------------
\subsubsection{Space-Time Diagram and Lattice of Snapshots}\label{sec:lattice of CGSs}

Global predicates are often specified over snapshots of the system. Under the asynchronous setting, the snapshot of a system is determined by the happen-before relation among events rather than by a global clock \cite{Babaoglu93, Schwarz94}. Specifically, a {\it global state} $\mathcal{G} = (s^{(1)}, s^{(2)}, \cdots, s^{(n)})$ is defined as a vector of local states from each $P^{(k)}$. If the constituent states of a global state $\mathcal{C}$ are pairwise concurrent, $\mathcal{C}$ is a {\it snapshot} (or consistent global state) \cite{Babaoglu93, Schwarz94}.

It is intuitive to define the {\it precede} relation (denoted by `$\prec$') between snapshots. Snapshots $\mathcal{C}_1 \prec \mathcal{C}_2$ iff $\mathcal{C}_2$ is obtained by advancing $\mathcal{C}_1$ on one process by one local state. The {\it lead-to} relation (denoted by `$\leadsto$') between snapshots is defined as the transitive closure of `$\prec$'. A {\it snapshot sequence} $\mathcal{S}(\mathcal{C}_i, \mathcal{C}_j)$ is a sequence of snapshots connected by `$\prec$'.

\begin{figure}[tbp]
  \centering
  \includegraphics[width=3.2in]{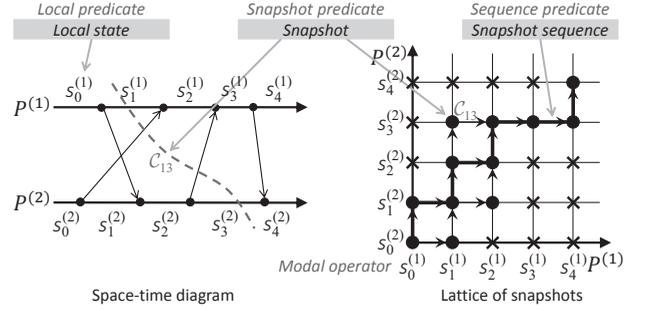}
  \caption{The space-time diagram and its corresponding lattice of snapshots.}
  \label{F:STD and lattice}
\end{figure}

The distributed computation can be intuitively illustrated by the {\it space-time diagram} ({\it STD}), e.g., in Fig. \ref{F:STD and lattice}. The STD contains multiple processes, and illustrates how the trace of the processes evolves over time. Global snapshots of the system can be depicted by the consistent cuts \cite{Schwarz94} over the STD. Snapshot sequences can be depicted by the consequent cuts over the STD.

One key notion in predicate detection is that the set of observed snapshots over the trace with the `$\leadsto$' relation has the {\it lattice} structure ({\it LAT}) \cite{Babaoglu93, Schwarz94}. For example, Fig. \ref{F:STD and lattice} shows the LAT of snapshots corresponding to the STD. The dots denote snapshots, crosses (`$\times$') denote inconsistent global states, and edges depict the `$\prec$' relation. The bold line is a possible snapshot sequence.

In the STD, we can see the detailed evolution of each process, but it is not convenient to reason over the global snapshots or snapshot sequences of the whole system. In comparison, global snapshots are dots in the LAT, and snapshot sequences are paths of consequent dots in the LAT. Though the LAT is more efficient to illustrate the global behavior of the system, it is usually computationally expensive to construct the LAT \cite{Cooper91, Schwarz94}. The STD and the LAT are different illustrations of the same computation, and should be used according to different needs.

Due to the uncertainty caused by the asynchrony, we can obtain multiple snapshot sequences from the LAT delineating all possible executions of the system. We only know that the actual execution of the system is one of these sequences but never know which one \cite{Schwarz94, Garg96, Babaoglu96, Yang13, Huang12, Wei12}.
%--

%------------------------------------------------------------------------
\subsection{Specification of Global Predicates}\label{sec:Specification}

Formal specification of predicates expresses the users' concerns on properties of the system. Formal specifications are precise and have the advantage of being amenable to automatic analysis and manipulation \cite{Wing90}. Based on the modeling of the system behavior discussed above, we can specify two classes of predicates: {\it snapshot predicates} and {\it sequence predicates} \cite{Yang13}. Snapshot predicates are properties that can be evaluated over a snapshot of the system, while sequence predicates are properties that can be evaluated over dynamic system behavior, i.e., the snapshot sequence, as shown in Fig. \ref{F:STD and lattice}. For example, properties $\phi_1$, $\phi_2$, and $\phi_3$ in Section \ref{sec:Case study} can be specified as snapshot predicates, and $\phi_4$ can be specified as a sequence predicate. In what follows, we briefly introduce these two classes of predicates.

%------------------------------------
\subsubsection{Snapshot Predicates}\label{sec:snapshot predicates}

Snapshot predicates can be evaluated based on the information with respect to one snapshot. We mainly focus on two types of snapshot predicates below.

{\it Conjunctive predicates} delineate the concurrency among local activities (which are delineated by local predicates) on each process. They are widely used, and can also serve as the basis for the specification of many other types of predicates \cite{Huang12}. Conjunctive predicates are defined as follows:
%--
\begin{eqnarray*}
 CP &::=& Pos(\phi)~|~Def(\phi) \\
 \phi &::=& \phi^{(1)}\wedge\phi^{(2)}\wedge\cdots\wedge\phi^{(n)} \\
 \phi^{(k)} &::=& \mbox{local predicate on } P^{(k)} (1\leq k\leq n)
\end{eqnarray*}

\noindent where the value of $\phi^{(k)}$ only depends on the local variable $x^{(k)}$ at $P^{(k)}$, and modal operators $Pos()$ and $Def()$ are adopted to cope with the multiple possible snapshot sequences resulting from the asynchrony discussed in Section \ref{sec:lattice of CGSs} \cite{Cooper91}. Informally, $Pos(\phi)$ means that predicate $\phi$ holds on one possible snapshot sequence, while $Def(\phi)$ means that predicate $\phi$ holds on all possible snapshot sequences. Detailed discussions on the semantics can be found in \cite{Huang12}.

Property $\phi_1$ in Section \ref{sec:Case study} can be expressed as a conjunctive predicate ``$Pos(\textrm{R1 detects the leak} \wedge \textrm{R2 detects the leak} \wedge \textrm{R3 detects the leak})$''.

The other important type of snapshot predicates is the {\it relational predicate}, which is defined as follows:
%--
\begin{eqnarray*}
 RP &::=& Pos(\phi)~|~Def(\phi) \\
 \phi &::=& x^{(1)} + x^{(2)} + \cdots + x^{(n)} \mbox{ \textit{relop} } C\\
 x^{(k)} &::=& \mbox{local variable at process } P^{(k)} (1\leq k\leq n)\\
 \mbox{\textit{relop}} &::=& < | > | \leq | \geq | =
\end{eqnarray*}

\noindent where $C$ is a constant. Detailed discussions on the semantics can be found in \cite{Mittal01}.

Property $\phi_2$ in Section \ref{sec:Case study} can be expressed as a relational predicate ``$Pos$(temperature\_R1 + temperature\_R2 + temperature\_R3 $>$ 3 $\times$ threshold)''.

%------------------------------------
\subsubsection{Sequence Predicates}\label{sec:sequence predicates}

Sequence predicates can be evaluated based on the information on a snapshot sequence. In this section, we present two important types of sequence predicates, {\it regular expression predicates} and {\it temporal logic predicates}, which can be employed to characterize behavioral patterns of the asynchronous computation.

Regular expression predicates can be specified as follows \cite{Yang13, Babaoglu96}:
%--
\begin{eqnarray*}
 REP &::=& Pos(\Phi)~|~Def(\Phi) \\
 \Phi &::=& \varnothing~|~\varepsilon~|~a~|~\Phi + \Phi~|~\Phi \cdot \Phi~|~\Phi^*
\end{eqnarray*}
\noindent where $\varnothing$ denotes the empty set, $\varepsilon$ denotes an empty word, and $a\in\Sigma$. The alphabet $\Sigma$ is the set of snapshot predicates involved in the regular expression predicates. Regular expression predicates unify a large body of important predicates including linked predicates \cite{Miller88}, interval-constrained sequence predicates \cite{Babaoglu95}, etc. Detection of regular expression predicates can boil down to a language recognition problem \cite{Yang13, Babaoglu96}. Detailed discussions on the semantics can be found in \cite{Yang13}.

Property $\phi_4$ in Section \ref{sec:Case study} can be expressed as a regular expression predicate ``$Def(a*ab*bc*cd*d)$'' with $a$ indicates that R1, R2, and R3 are all in workshop A, $b$ indicates that R2 and R3 are in workshop A while R1 is in workshop B, $c$ indicates that R3 is in workshop A while R1 and R2 are in workshop B, and $d$ indicates that R1, R2, and R3 are all in workshop B.

An important type of temporal logic predicates can be specified as CTL formulae, which can effectively express the notion of branching time. CTL has a two-stage syntax where formulae in CTL are classified into state and path formulae \cite{Wei12}. The state formulae are assertions about the atomic propositions (i.e., snapshot predicates) in the states (i.e., snapshots) and their branching structure, while path formulae express temporal properties of paths (i.e., snapshot sequences).

CTL state formulae over the set $\Sigma$ of atomic propositions are formed according to the following grammar:
$$\Phi ::= \top \mid a \mid \Phi_1 \land \Phi_2  \mid \neg \Phi \mid \exists \varphi \mid \forall \varphi$$
where $a \in \Sigma$ and $\varphi$ is a path formula. CTL path formulae are formed according to the following grammar:
$$\varphi ::= \bigcirc \Phi \mid \Phi_1 \pmb \cup \Phi_2$$
where $\Phi$, $\Phi_1$, and $\Phi_2$ are state formulae. Detailed discussions on the semantics can be found in \cite{Wei12}.

%------------------------------------------------------------------------
\subsection{Detection of Specified Predicates}\label{sec:Detection}

The predicate detection algorithms often need to be online, which is decided by the application scenarios. For example in a context-aware computing scenario, the application has close interaction with the computing environment and needs to get informed as soon as possible when the predicate of its concern turns true. We design online predicate detection algorithms for such scenarios.

The detection algorithms also need to be incremental, i.e., only the newly updated part of the trace of the distributed system is checked. The trace of the distributed system continuously evolves as the system runs. Each time the trace evolves (i.e., a new local state arrives), the checker process $P_{che}$ should check whether the new part of the trace changes the value of the specified predicate.

In principle, we can design a general algorithm to detect all types of predicates defined in Section \ref{sec:Specification}. However, the general algorithm is usually prohibitively expensive \cite{Cooper91}. Thus we design specific efficient (but not general) algorithms for different types of predicates \cite{Huang12, Yang13, Wei12}.

%========================================================================
\section{The PD-CA Conceptual Framework}\label{sec:PD approach}

In this section, we first describe the essential concepts in the Predicate-Detection-based Context-Awareness (PD-CA) framework. Then we discuss how these concepts facilitate the development of context-aware applications.

A context-aware computing scenario adopts a middleware-based architecture \cite{Xu05, Xu10, Huang12, Yang13}. Context collecting devices send local contexts to the context-aware middleware. The middleware processes the contexts, evaluates the contextual properties specified by the application, and notifies the application to conduct context-aware behavior accordingly. This architecture will be detailed in Section \ref{sec:Architecture and implementation} and \ref{sec:develop app}.

The PD-CA framework solves the challenges in Section \ref{sec:Challenges} and enables context-awareness based on the predicate detection theory. The mapping of concepts between the predicate detection theory and the PD-CA framework is listed in Table \ref{T:PD and SA}.
%--
\begin{table}[tbp]
    \centering
        \scriptsize{
        \begin{tabular}{|p{1.26cm}|p{1.96cm}|p{3.96cm}|}
          \hline
          & Concepts in PD & Concepts in PD-CA \\\hline\hline
          \multirow{11}*{Modeling}
            & \tabincell{l}{instrumented\\ process} & \tabincell{l}{context collecting process on the\\ context collecting device} \\ \cline{2-3}
            & \tabincell{l}{local variable} & \tabincell{l}{type of local context provided by\\ the context collecting device} \\ \cline{2-3}
            & \tabincell{l}{local event} & \tabincell{l}{contextual event: value change of\\ local variable/predicate} \\ \cline{2-3}
            & \tabincell{l}{local state} & \tabincell{l}{local context (value of local variable\\/predicate) with logical timestamp} \\ \cline{2-3}
            & \tabincell{l}{global snapshot} & \tabincell{l}{global snapshot of the system of\\ context collecting devices} \\ \cline{2-3}
            & \tabincell{l}{snapshot sequence} & \tabincell{l}{dynamic behavior of the system of\\ context collecting devices} \\ \hline\hline
          \multirow{5}*{Specification}
            & \tabincell{l}{local predicate} & \tabincell{l}{local contextual property of a\\ context collecting device} \\ \cline{2-3}
            & \tabincell{l}{snapshot predicate} & \tabincell{l}{global instantaneous property of the\\ system of context collecting devices} \\ \cline{2-3}
            & \tabincell{l}{sequence predicate} & \tabincell{l}{dynamic behavioral property of the\\ system of context collecting devices} \\ \hline\hline
          \multirow{3}*{Detection}
            & \tabincell{l}{checker process} & \tabincell{l}{property detection process on the\\ property detection server} \\ \cline{2-3}
            & \tabincell{l}{detection of\\ predicates} & \tabincell{l}{persistent monitoring of contextual\\ properties} \\ \hline
        \end{tabular}
    }
  \caption{Enabling context-awareness based on predicate detection}
  \label{T:PD and SA}
\end{table}

%------------------------------------------------------------------------
\subsection{Modeling of Asynchronous Context}\label{sec:Modeling of PD-CA}

The key problem in the modeling of asynchronous context is to cope with the asynchrony originating from the mismatching between the capability-constrained context collecting devices and the rich semantics of the computing environment. The contexts are spatially distributed and temporally asynchronous. We discuss these two characteristics in detail.

%------------------------------------
\subsubsection{Modeling the Spatially Distributed Context}\label{sec:Modeling distributed context}

As discussed in Section \ref{sec:Case study}, a context-aware application is often concerned with and reacts to the global properties of its contexts. While one single context collecting device can only provide limited types of contexts, the contexts of an application are often provided by a group of coordinating context collecting devices. These context collecting devices persistently collect local contexts and send them to the context-aware middleware. The middleware then combines these local contexts into global contexts for the detection of global properties specified by the application.

In our PD-CA framework, the coordinating context collecting devices are modeled as {\it context collecting processes}, and form an {\it asynchronous system of context collecting processes}. The devices coordinate by exchanging messages. The {\it type of local context} provided by a device is modeled as a local variable. The value of local context and its logical timestamp define a local state of the device. Value changes of local contexts are modeled as {\it contextual events}.

Based on our intuitive understanding of the computing environment, we model the spatially distributed context as a hierarchy of global-local context. {\it Local contexts} are defined on local states of context collecting devices, while {\it global contexts} are defined on global system snapshots (i.e., vectors of local states of each context collecting device). In our PD-CA framework, the hierarchy of specification is based on the hierarchy of global-local context, as detailed in Section \ref{sec:Specification of PD-CA}.

%------------------------------------
\subsubsection{Modeling the Temporally Asynchronous Context}\label{sec:Modeling asynchronous context}

As discussed in Section \ref{sec:Motivation and challenges}, the distributed context collecting devices coordinate in a loosely-coupled asynchronous manner. Thus the local contexts collected by them are intrinsically asynchronous. Without global clocks, we cannot easily tell whether a group of local contexts collected from different devices forms a meaningful global context.

In our PD-CA framework, due to the asynchrony, we do not rely on global clocks, but on logical time to model the temporal order among the local states of different context collecting devices. The temporal order is defined by the happen-before relation resulting from message exchanges between the context collecting devices (defined in Section \ref{sec:asynchronous message-passing}). Local states from different context collecting devices, which have no happen-before relation among each other, form a meaningful snapshot, as introduced in Section \ref{sec:lattice of CGSs}. The local contexts of a meaningful snapshot form a meaningful global context. As the context collecting devices persistently collect contexts, we can get sequences of meaningful snapshots. Since the happen-before relation is a partial order relation, we can observe multiple possible snapshot sequences, but cannot tell which is the actual one. This uncertainty caused by the asynchrony is explicitly modeled in the STD or the LAT (in Section \ref{sec:lattice of CGSs}).

%------------------------------------------------------------------------
\subsection{Specification of Contextual Properties}\label{sec:Specification of PD-CA}

In our PD-CA framework, the context-aware application expresses its concerns on contextual properties of the computing environment by specification of (usually global) predicates. The hierarchical specification of predicates is directly based on the hierarchy of global-local context, as discussed in Section \ref{sec:Modeling distributed context}.

%------------------------------------
\subsubsection{The Hierarchy of Predicates}\label{sec:Hierarchy of predicates}

We introduce the hierarchy of predicate specification in a bottom-up manner, and illustrate it in Fig. \ref{F:STD and lattice}.
%--
\begin{itemize}
  \item \textit{Local predicates}. Local predicates are specified over local states of context collecting devices, to indicate the application's concern on specific aspects of the computing environment. For example, ``R1 detects the leak of hazardous material'' (in property $\phi_1$ in Section \ref{sec:Case study}) is a local predicate.
  \item \textit{Snapshot predicates}. Snapshot predicates are specified over global snapshots (as discussed in Section \ref{sec:snapshot predicates}), to indicate the application's concern on global instantaneous properties of the computing environment. For example, ``the average temperature is greater than the threshold'' (property $\phi_2$ in Section \ref{sec:Case study}) is a snapshot predicate.
  \item \textit{Sequence predicates}. Based on the modeling of temporal evolution of context, the application can specify sequence predicates to describe the dynamic properties and behavioral patterns of the computing environment on snapshot sequences. For example, ``the robots pass the gateway in certain order'' (property $\phi_4$ in Section \ref{sec:Case study}) is a sequence predicate.
  \item \textit{Modal operators}. To cope with the uncertainty resulting from the asynchrony, modal operators $Pos()$ and $Def()$ are adopted to interpret the predicates over the multiple possible snapshot sequences, as discussed in Section \ref{sec:snapshot predicates}. Each of the two modal operators has its own characteristic. $Pos()$ is often used to eliminate the possibility of ``bad things'', while $Def()$ is often used to ensure the occurrence of ``good things''. For example, we use the modal operator $Pos()$ in property $\phi_2$ (formalized in Section \ref{sec:snapshot predicates}), since we need to prevent fire accidents in all possible snapshot sequences.
\end{itemize}

%------------------------------------
\subsubsection{Separation of Concerns}\label{sec:Separation of concern}

\begin{figure}[tbp]
  \centering
  \includegraphics[width=3.2in]{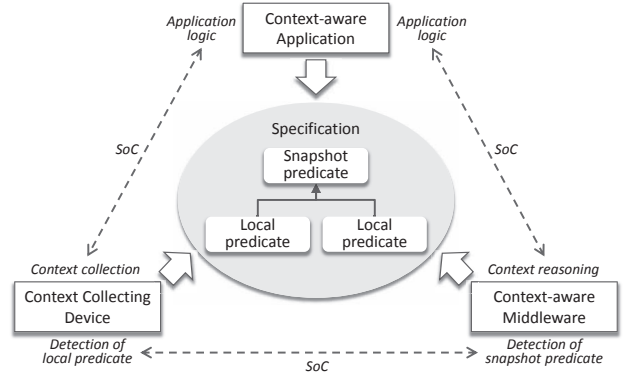}
  \caption{Separation of Concerns (SoC).}
  \label{F:Separate of concern}
\end{figure}

The specification of contextual properties greatly facilitates the separation of concerns among the application, the context collecting devices, and the middleware, as shown in Fig. \ref{F:Separate of concern}.

Contextual properties specified in predicates provide a logical abstraction of the computing environment (i.e., the asynchronous system of context collecting devices) and hide the complexity of low-level programming to the application. In this way, the application only have to specify predicates to the middleware and react to the satisfaction of the predicates, not having to know how the asynchronous contexts are collected and processed. The middleware collects the related asynchronous contexts, detects the specified predicates, and notifies the application when the predicates turn true.

As the specified global property is often in the form of a hierarchy of snapshot-local predicates, the middleware can subscribe consisting local predicates to the corresponding context collecting devices. Rather than sending all the collected low-level contexts (values of each local variable $x^{(k)}$ in our PD-CA framework) to the middleware, each context collecting device detects the subscribed local predicate, and sends the filtered local contexts (values of the local predicate) to the middleware when the value of the predicate changes. The middleware processes the aggregated local contexts and detects the specified predicate under the guidance of the PD-CA framework. It does not have to know how the local contexts are collected and how the local predicates are evaluated.
%--

%------------------------------------------------------------------------
\subsection{Detection of Specified Properties}\label{sec:Detection of PD-CA}

In the detection of specified properties, the checker process is instantiated as a {\it property detection process} deployed on the middleware, which detects the specified predicates in an online and incremental manner.

Since context-awareness is such a persistent process that new emerging contexts trigger the application to conduct context-aware behavior, our detection of specified predicates is achieved in an event-driven manner. When new contexts are generated by a context collecting device, they will first be filtered by a local predicate, modeled as contextual events (indicating the value changes of the local predicate), and sent to the corresponding property detection process on the middleware. When receiving a contextual event, the property detection process detects whether the newly arrived context makes the specified predicate true and notifies the application accordingly.

Though specific predicate detection algorithms are designed for different types of predicates (as discussed in Section \ref{sec:Detection}), predicate detection in our PD-CA framework is achieved in two steps generally. The property detection process first maintains the STD or the LAT of global contexts, and then invokes the corresponding predicate detection algorithm to detect the specified predicate over the STD or the LAT. The maintenance of the STD or the LAT is generally reusable and independent to the detection of specific types of predicates. Implementation details of the predicate detection algorithms will be discussed in Section \ref{sec:Property detection layer}.

%------------------------------------------------------------------------
\subsection{Design Process of PD-CA}\label{sec:Design process}

After discussing essentials of the PD-CA framework, we further discuss the process of developing context-aware applications using the framework. Based on predicate detection, context-awareness in asynchronous environments is achieved by runtime detection of contextual properties. The middleware receives the contextual property from the application, persistently collects contexts from multiple distributed context collecting devices, detects whether the property holds, and notifies the application accordingly. Specifically, the process of developing a context-aware application using the PD-CA framework includes five steps, as detailed below.

%--
\textit{Step 1. Obtainment of the specification.}
The application specifies the contextual property to the middleware. The middleware then parses the property, identifies the consisting local predicates in the property, finds the related context collecting devices, registers the local predicates to their corresponding context collecting devices, and launches the property detection process which is in charge of detecting the contextual property.

%--
\textit{Step 2. Acquisition of the asynchronous context.}
Each context collecting device registers itself to the middleware with the type of context it provides. When receiving the registration of a local predicate, the context collecting process on the context collecting device filters local collected contexts with the local predicate and generates local states. When the context collecting devices exchange messages between each other, the logical vector clocks are updated by piggybacking logical timestamps on messages \cite{Huang12,Yang13}. When a context collecting device sends/receives messages or the value of the local predicate changes, the context collecting process sends the local state (local context with current logical timestamp) to the property detection process on the middleware.

%--
\textit{Step 3. Processing of the asynchronous context.}
When the property detection process on the middleware receives the local contexts from distributed context collecting processes, it first explicitly maintains the STD or the LAT based on the causal relation among these asynchronous contexts, and then detects the specified contextual property.

%--
\textit{Step 4. Notifying the context-aware application.}
If the contextual property is evaluated true, the middleware notifies the context-aware application of the satisfaction to conduct context-aware behavior.

%--
\textit{Step 5. Taking context-aware behavior.}
The context-aware behavior of the application is constructed around the contextual property. When notified of the satisfaction of the property, the application conducts the corresponding context-aware behavior.

By providing middleware support for this design process, the development of context-aware applications can be greatly simplified. The application developers can be relieved of the time- and energy-consuming burden of collecting asynchronous contexts from distributed heterogeneous context collecting devices and processing these asynchronous contexts, as in Step 2-4. The application developers only need to specify contextual properties and implement the corresponding context-aware behavior.

In the following Section \ref{sec:Architecture and implementation} and \ref{sec:develop app}, we discuss the design and implementation of the middleware and the application, respectively.

%========================================================================
\section{MIPA - Providing Middleware Support for the PD-CA Framework}\label{sec:Architecture and implementation}

In this section, we present the design and implementation of MIPA -- {\it Middleware Infrastructure for Predicate detection in Asynchronous environments}, which would greatly simplify the tasks of building context-aware applications with the PD-CA framework \cite{Huang12, Yang13, Wei12, Huang10b, MIPA}.

The design process of the PD-CA framework (in Section \ref{sec:Design process}) consists of several logically independent steps. Thus we adopt a layered architecture for MIPA. The layered architecture groups related functionalities into distinct layers and provides software engineering benefits such as separation of concerns, information hiding, extensibility, and reusability.
%--
\begin{figure}[tbp]
  \centering
  \includegraphics[width=3.2in]{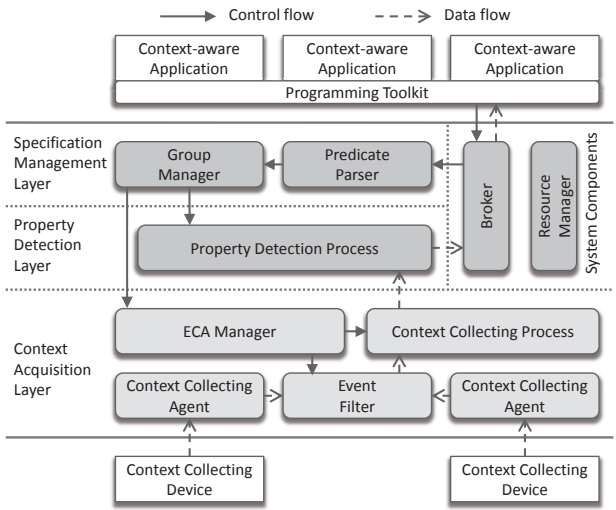}
  \caption{The architecture of MIPA.}
  \label{F:Architecture}
\end{figure}

The architecture of MIPA is listed in Fig. \ref{F:Architecture}. The middleware can be deployed as a centralized property detection server (components in dark grey color), and several context collecting clients (components in light grey color) over context collecting devices. The main functions of each layer are explained below:
%--
\begin{itemize}
    \item The {\it specification management layer} is in charge of obtaining the contextual property from the application, locating the corresponding context collecting devices, and maintaining the group of processes involved in the detection of the property.
    \item The {\it context acquisition layer} is in charge of collecting asynchronous contexts on context collecting devices and sending them to the property detection server for predicate detection.
    \item The {\it property detection layer} receives the local contexts sent from the context collecting devices and detects the specified contextual property. When the contextual property is detected true, the property detection layer will notify the application.
\end{itemize}

\noindent We first overview the system components. Then, we discuss the design and implementation of each layer.

%------------------------------------------------------------------------
\subsection{System Components}\label{sec:System components}

To support the components involved in the processing of asynchronous contexts and ensure a strong decoupling among them, several system components are provided.

%------------------------------------
\subsubsection{Broker}

\begin{figure}[tbp]
  \centering
  \includegraphics[width=3.2in]{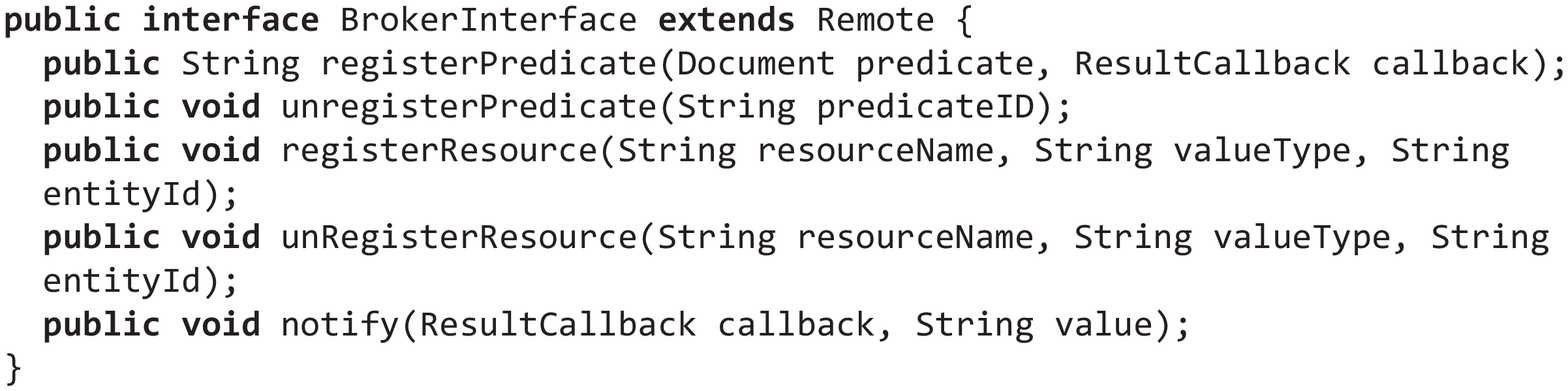}
  \caption{The Broker of MIPA.}
  \label{F:Broker}
\end{figure}

The {\it broker} is in charge of interacting with the application and context collecting clients, and delivering their requests to the corresponding components, as shown in Fig. \ref{F:Broker}. Specifically, the application registers the contextual property to the middleware through the broker and the broker delivers the property to the specification management layer for further processing. When the application unregisters the property, the broker notifies the specification management layer to stop the processes involved in the detection of the property. When the property is detected true, the middleware notifies the application through the broker. Similarly, the context collecting devices contact the broker to register/unregister themselves on the middleware.

%------------------------------------
\subsubsection{Resource Manager}

The {\it resource manager} is in charge of the management of the available types of contexts and their corresponding context collecting devices. It supports the registration/unregistration, as well as the lookup of context collecting devices.

When a context collecting device registers to the middleware (through the broker), it will register its location and the type of context it provides to the resource manager. When specifying contextual properties, the application might first query the resource manager to see which types of contexts are available in the context-aware computing scenario, and then specify contextual properties based on the available types of contexts. When receiving a contextual property from the application, the specification management layer would extract the consisting types of contexts and query the resource manager for the locations of the corresponding context collecting devices.
%--

%------------------------------------------------------------------------
\subsection{Specification Management Layer}\label{sec:Specification layer}

The {\it specification management layer} is in charge of the management of the specification and the processes involved in the detection of the specification. When a contextual property is received from the application through the broker, it should first be parsed and the related context collecting devices should be identified. Then, the corresponding processes, including the context collecting processes on the context collecting clients and the property detection process on the property detection server, are launched for the detection of the property. We detail the management of the specification and the group of processes involved in the predicate detection below.

%------------------------------------
\subsubsection{Management of the Specification}\label{sec:Management of the specification}

\begin{figure}[tbp]
  \centering
  \includegraphics[width=3.2in]{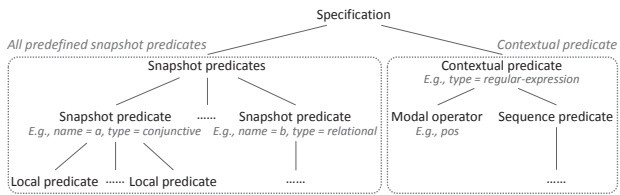}
  \caption{The tree structure of the specification.}
  \label{F:specification structure}
\end{figure}

The specification serves as the contract between the context-aware application and the context-aware middleware \cite{Wing90}. The predicates may have different types, such as conjunctive predicates and regular expression predicates (in Section \ref{sec:Specification}). We provide a uniform mechanism to represent different types of predicates. Since sequence predicates are generally composed of several snapshot predicates, we extract these snapshot predicates as an alphabet to facilitate the description of sequence predicates. As the snapshot predicates are often in the form of hierarchies of snapshot-local predicates, our specification mechanism also adopts a hierarchical structure, as shown in Fig. \ref{F:specification structure}.
%--

Specifically, the ``{\it Snapshot predicates}'' node contains all consisting snapshot predicates as the alphabet, and each ``{\it Snapshot predicate}'' node defines a snapshot predicate. The ``{\it Contextual predicate}'' node contains the specified predicate which is composed of snapshot predicates defined in the ``{\it Snapshot predicate}'' nodes.\footnote{Snapshot predicates can be viewed as sequence predicates with only one snapshot predicate.} Each ``{\it Snapshot predicate}'' node is labeled with its type, such as \textit{conjunctive} or \textit{relational}, as introduced in Section \ref{sec:snapshot predicates}. The ``{\it Contextual predicate}'' node is also labeled with its type, such as \textit{regular-expression} or \textit{CTL}, as introduced in Section \ref{sec:sequence predicates}. This structure can unify a large amount of typical predicates as introduced in Section \ref{sec:Specification}, and is easy to extend new types of snapshot or sequence predicates.

\begin{figure}[tbp]
  \centering
  \includegraphics[width=3.2in]{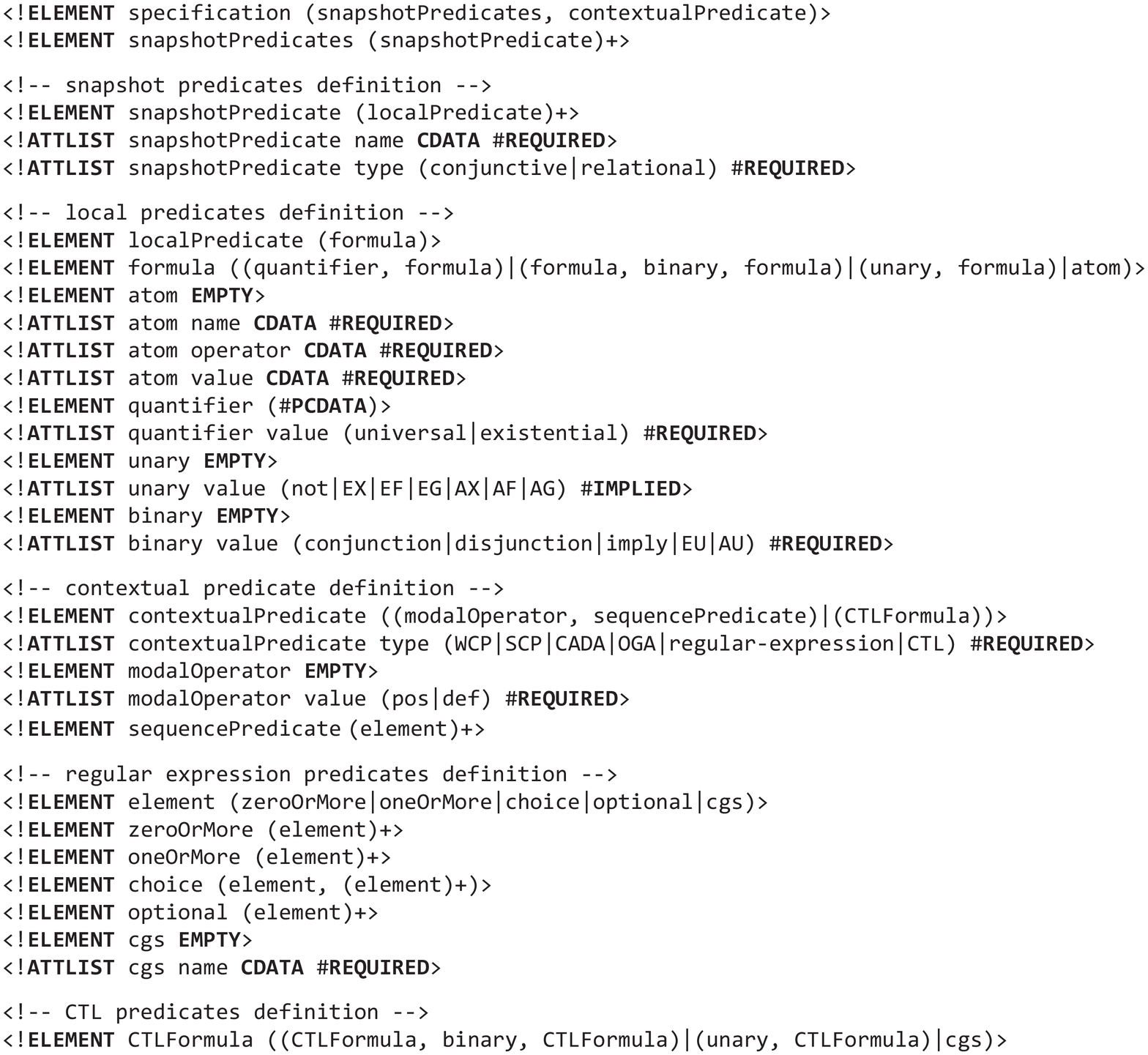}
  \caption{The DTD of the specification.}
  \label{F:DTD of specification}
\end{figure}

\begin{figure}[tbp]
  \centering
  \includegraphics[width=3.2in]{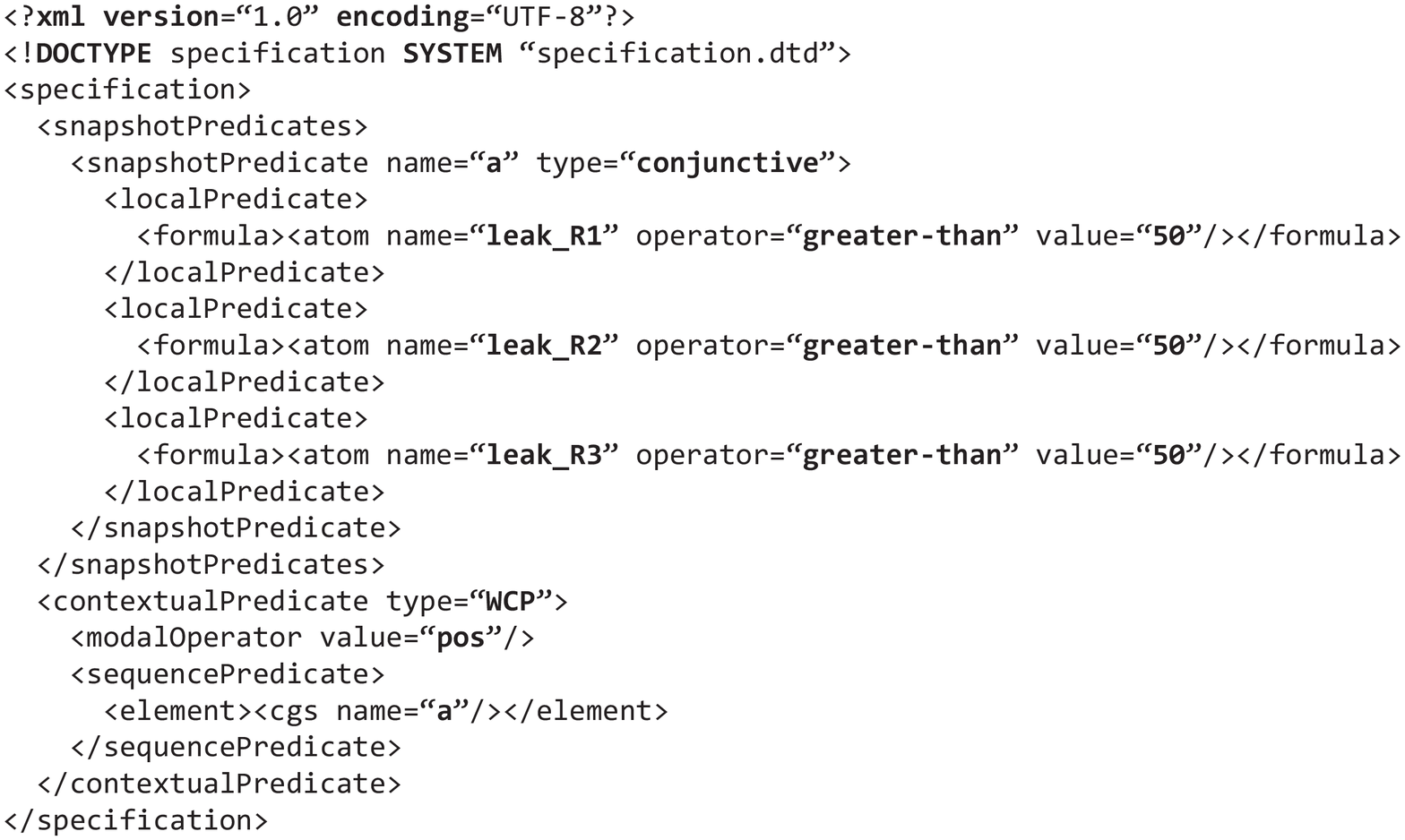}
  \caption{Property $\phi_1$ in XML.}
  \label{F:specification example}
\end{figure}

We adopt XML which is a flexible way to create ``self-describing data'' and naturally has the hierarchical structure to realize our specification mechanism. The DTD of the specification is listed in Fig. \ref{F:DTD of specification}. It currently supports conjunctive predicates, regular expression predicates, and CTL predicates. New types of predicates can be added by extending the ``{\it snapshotPredicate}'' and ``{\it contextualPredicate}'' in Fig. \ref{F:DTD of specification}. As an example, property $\phi_1$ in Section \ref{sec:Case study}, which is formalized in Section \ref{sec:snapshot predicates}, is shown in Fig. \ref{F:specification example}.
%--
%--

When receiving the specification in XML, a {\it predicate parser} is employed to parse the specification into the structure shown in Fig. \ref{F:specification structure}. Notice that different types of snapshot and sequence predicates have different structures. Thus, we provide different parsers for different types of predicates. For example, when the predicate parser reaches a snapshot predicate labeled with ``{\it conjunctive}'' as shown in Fig. \ref{F:specification structure}, the type is first recognized and the corresponding parser for conjunctive predicates is invoked to parse the predicate. Our design provides good reusability and extensibility. New types of predicates can be parsed by adding new parsers accordingly.

According to the type of the contextual predicate, the corresponding processes which are dedicated to the detection of the predicate can be further launched.

%------------------------------------
\subsubsection{Management of the Group of Processes Involved in the Predicate Detection}

After parsing the predicate, the consisting local predicates are identified. The middleware locates the context collecting devices corresponding to the local predicates through the resource manager, registers the local predicates to these devices (context collecting clients) to create context collecting processes which detect the local predicates, and creates a property detection process which detects the specified predicate on the property detection server.

These processes are distributed on different context collecting clients and the property detection server. To better maintain them, we organize the processes created for a predicate as a {\it predicate detection group}, and adopt a {\it group manager} to maintain the groups.

Specifically, when a predicate is registered/unregistered, the group of corresponding processes is created/destroyed. Notice that the context collecting processes may send messages between each other or to the property detection process. Thus, the property detection process should be launched before the context collecting processes, and the context collecting processes should not be started until they are all created. Likewise, during destruction, the context collecting processes should be destroyed before the property detection process. The creation and destruction are coordinated by the group manager.

%------------------------------------------------------------------------
\subsection{Context Acquisition Layer}\label{sec:Context acquisition layer}

The {\it context acquisition layer} manages the registration/unregistration of context collecting devices, and acquires local contexts from them.

%------------------------------------
\subsubsection{Management of Context Collecting Devices}

With the development of embedded computing technologies, various new C$^3$S devices are developed to support context-awareness. These devices register themselves to the middleware, coordinate to collect local contexts, and send local contexts to the middleware. Local contexts include information about the {\it physical environment} such as hazardous material leak and indoor temperature, as well as the {\it logical environment} delineating the status of coordination among the context collecting devices, such as the neighbor information.

To shield the application from low-level details of heterogeneous devices, all providers of all types of contexts are abstracted as {\it context providers} by implementing the same interface {\sf ContextProvider}. Each device registers the types of local contexts it provides to the middleware by registering the corresponding context providers to the middleware. Therefore, the middleware can treat different types of local contexts from different devices in a similar way, which eases further processing of local contexts. To each context provider, a {\it context collecting agent} is assigned. These context collecting agents are in charge of manipulating the context providers and sending collected data from context providers to the context collecting processes.

\begin{figure}[tbp]
  \centering
  \includegraphics[width=2.6in]{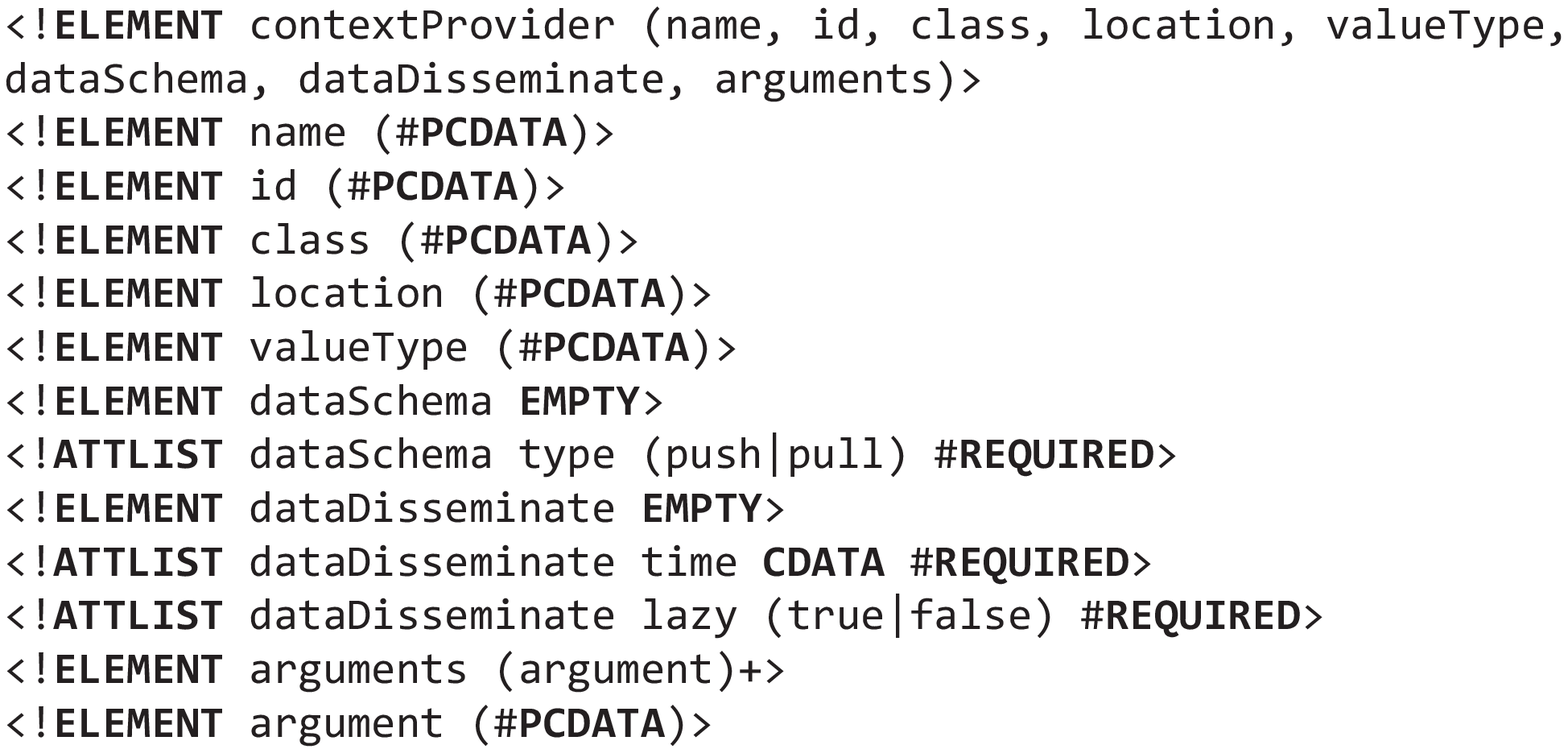}
  \caption{The DTD of the context provider configuration.}
  \label{F:DTD of sensors}
\end{figure}

\begin{figure}[tbp]
  \centering
  \includegraphics[width=2.6in]{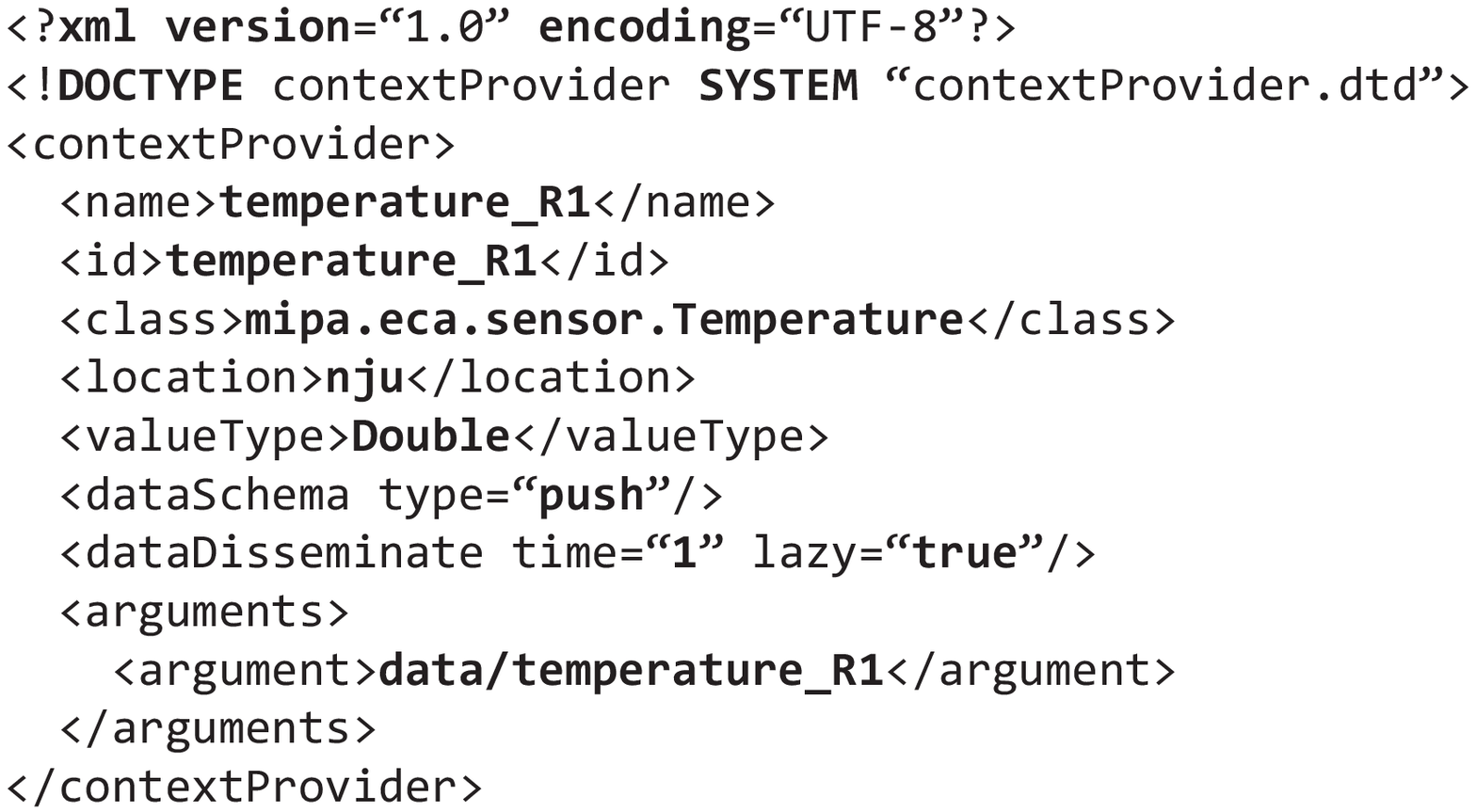}
  \caption{The context provider configuration for a temperature sensor of robot R1.}
  \label{F:XML cfg of sensor}
\end{figure}

To ease the integration of new context collecting devices, context collecting agents are implemented as plug-ins. When registering to the middleware, new devices register each context provider to the middleware using a {\it context provider configuration} in XML, i.e., a document containing the local context it provides and all the information needed to interact with the hardware or software modules of the device \cite{Schreiber12, Aberer06}. The DTD of the XML for context provider configuration is listed in Fig. \ref{F:DTD of sensors}, and an exemplar configuration of a temperature sensor is listed in Fig. \ref{F:XML cfg of sensor}. As shown in Fig. \ref{F:XML cfg of sensor}, the temperature sensor provides the type of local context ``temperature\_R1''.

An {\it ECA manager} (introduced in Section \ref{sec:Acquisition of local context} in detail) on each context collecting client is employed to manage the registration/unregistration of the context collecting agents. When a new device registers to the middleware, the ECA manager on the device parses all the local context provider configurations, and assembles the corresponding context providers (e.g., the class {\sf mipa.eca.sensor.Temperature} in Fig. \ref{F:XML cfg of sensor} for the temperature sensor of robot R1) and context collecting agents to interact with the context providers. The ECA manager then registers the types of local contexts to the resource manager on the property detection server (through the broker).
%--
%--

%------------------------------------
\subsubsection{Acquisition of Local Contexts}\label{sec:Acquisition of local context}

\begin{figure}[tbp]
  \centering
  \includegraphics[width=3.2in]{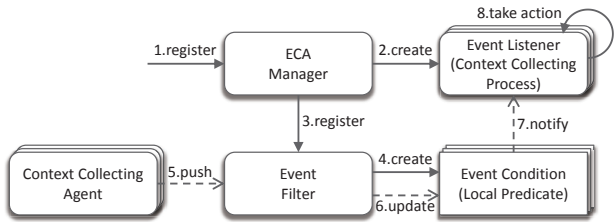}
  \caption{The ECA mechanism.}
  \label{F:ECA}
\end{figure}

To react to dynamic changes of the contexts, MIPA provides an efficient way to monitor the changes of contexts and to disseminate the changes to interested entities. As discussed in Section \ref{sec:Separation of concern}, the snapshot predicates often adopt the form of hierarchies of snapshot-local predicates, and the context collecting processes are only interested in the value changes of local predicates. We adopt the {\it Event-Condition-Action} (ECA) mechanism \cite{Paton99, Chan03, Xu05} to achieve persistent detection of local predicates, as shown in Fig. \ref{F:ECA}.

Specifically, when creating the group of processes dedicated to the detection of the specified predicate, the group manager locates the context collecting clients which provide the types of contexts required by the consisting local predicates, and registers each local predicate to the ECA manager of the corresponding context collecting client, as shown in Step 1 in Fig. \ref{F:ECA}. Then, the ECA manager creates an {\it event listener} (i.e., the {\it context collecting process}) which listens to the value changes of the local predicate, and registers the local predicate to the {\it event filter}, as in Step 2-3. In Step 4, the event filter takes the local predicate as the {\it event condition} on the corresponding context collecting agents. When the event filter receives the contexts pushed from the context collecting agents, it updates the value of the local predicate. When the value changes, the event filter notifies the context collecting process to take actions, as in Step 5-8.\footnote{In case of snapshot predicates which cannot be decomposed into a series of local predicates (e.g., the relational predicates), an empty condition is registered to the event filter, and all the contexts of the related context collecting agent will be pushed to the context collecting process without filtering.} We support local predicates in first-order logic composed of atomic predicates defined on local contexts.
%--

\begin{figure}[tbp]
  \centering
  \includegraphics[width=3.2in]{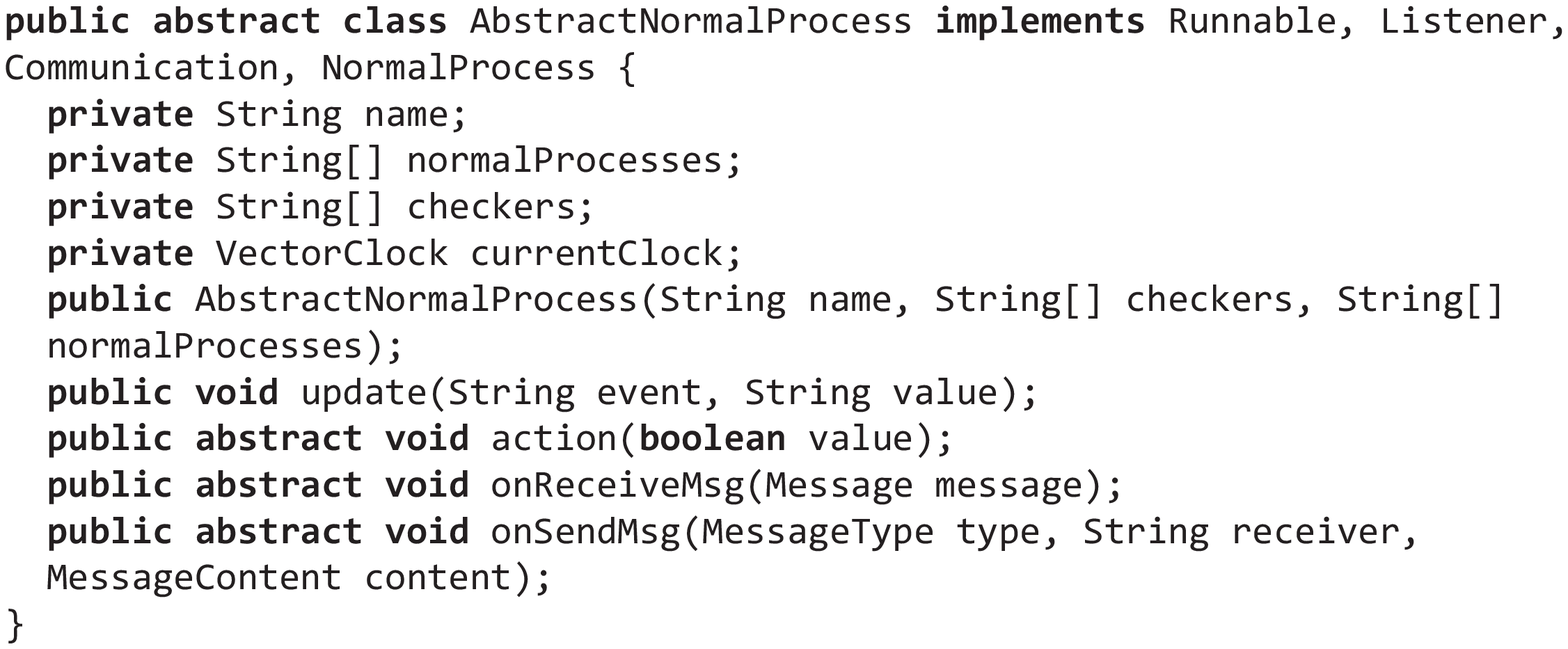}
  \caption{The abstract class AbstractNormalProcess.}
  \label{F:AbstractNP}
\end{figure}

The {\it context collecting process} is in charge of taking actions to event changes, including updating local logical vector clock and sending local contexts with logical timestamps to the property detection process on the property detection server. Notice that the detection algorithms for different types of predicates may have different logics of context collecting processes \cite{Huang12, Yang13, Wei12, Garg96}, but they share the same skeleton. Thus, we adopt the {\it template method} pattern \cite{Gamma95} to facilitate the development of new types of context collecting processes. New context collecting processes integrate to the middleware by extending the abstract class {\sf AbstractNormalProcess}, as shown in Fig. \ref{F:AbstractNP}.

In method {\sf action()} (i.e., Step 8 in Fig. \ref{F:ECA}), the concrete context collecting process receives the value changes of the local predicate, and it should send the local context with the current logical timestamp ({\sf currentClock} in Fig. \ref{F:AbstractNP}) to the property detection process. The methods {\sf onSendMsg()} and {\sf onReceiveMsg()} will be invoked when the local coordination algorithm on the device sends/receives messages to/from other context collecting devices.\footnote{This is achieved by registering the concrete context collecting process to the local coordination algorithm when the ECA manager creates the context collecting process.} Thus, the concrete context collecting process should update the logical vector clock in these two methods, to capture the happen-before relation between the local contexts on different context collecting devices, as required in Step 2 of the design process in Section \ref{sec:Design process}.

%------------------------------------------------------------------------
\subsection{Property Detection Layer}\label{sec:Property detection layer}
%--

The {\it property detection layer} detects the specified predicates and notifies the application of the satisfaction of the predicates.

As discussed in Section \ref{sec:Detection of PD-CA}, the detection is driven by the messages (containing the local contexts) sent from context collecting processes. When receiving the local contexts, the property detection process on the property detection server first maintains the STD or the LAT according to different requirements of the predicate detection algorithms, and then detects the specified predicate \cite{Huang12, Yang13, Wei12}. When the predicate is detected true, the property detection process will notify the application (through the broker).

The maintenance of the STD or the LAT is generally reusable and independent to the detection. Thus, this layer should support the maintenance of the STD or the LAT, and provide good extensibility for different types of property detection processes. We provide two abstract classes which achieve the maintenance of the STD and the LAT, respectively. Like the design of context collecting processes in Section \ref{sec:Acquisition of local context}, we also adopt the {\it template method} pattern \cite{Gamma95} to encapsulate different predicate detection algorithms and to facilitate the development of new property detection processes. We adopt a {\it checker factory} to manage the creation of property detection processes.
%--
\begin{figure}[tbp]
  \centering
  \includegraphics[width=3.2in]{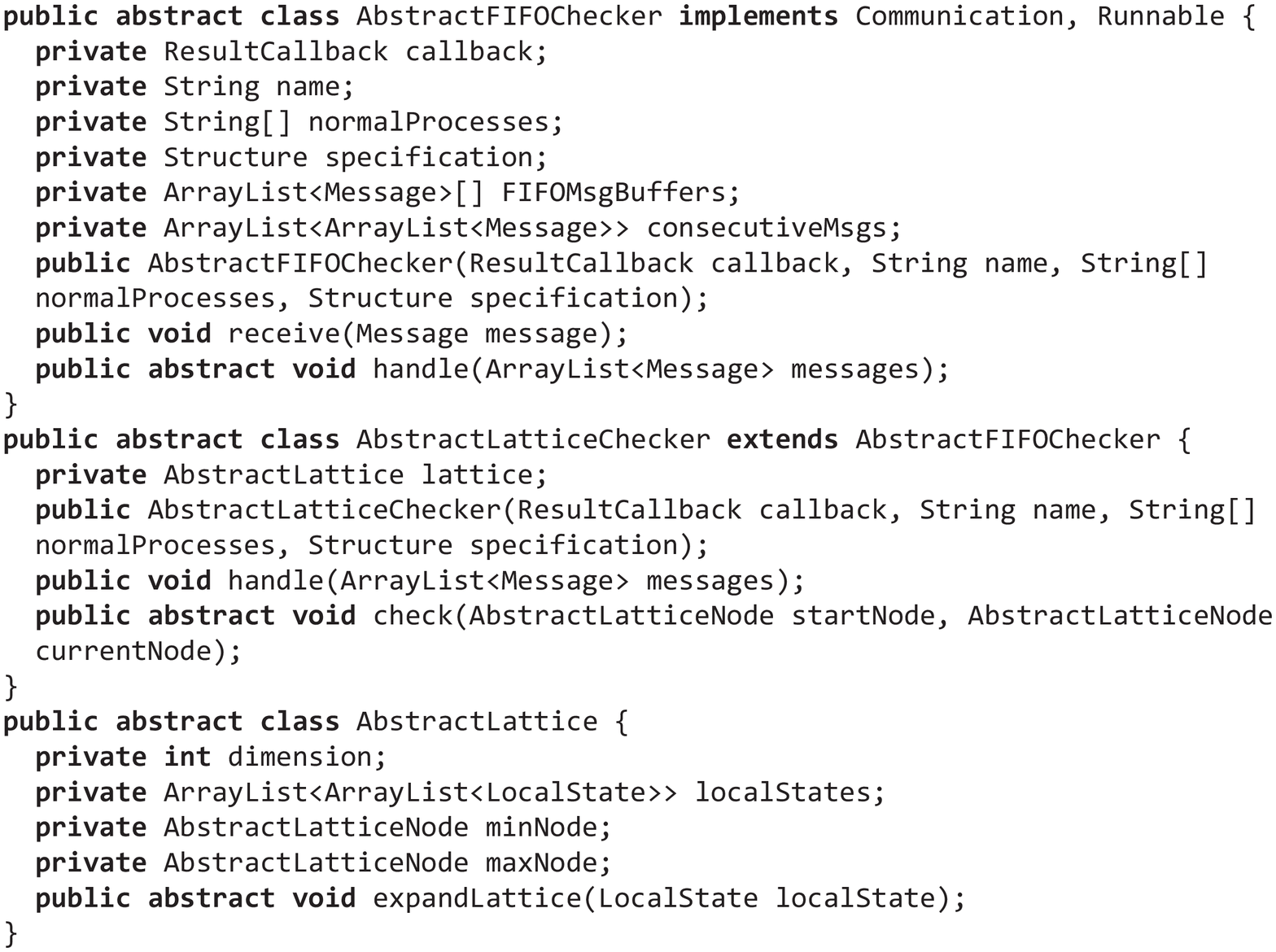}
  \caption{The abstract classes AbstractFIFOChecker, AbstractLatticeChecker, and AbstractLattice.}
  \label{F:AbstractChecker}
\end{figure}

%------------------------------------
\subsubsection{Property Detection over the STD}

The abstract class {\sf AbstractFIFOChecker} achieves the maintenance of the STD. Property detection processes which rely on the STD (e.g., \cite{Garg94, Garg96, Huang09, Huang12}), only need to extend the abstract class and implement the specific detection algorithms in the method {\sf handle()}, as shown in Fig. \ref{F:AbstractChecker}.

The {\sf AbstractFIFOChecker} adopts message buffers (the {\sf FIFOMsgBuffers}) to ensure that messages from the same context collecting process are FIFO. The maintenance of the STD is transparent to the detection algorithms, which relieves the detection algorithms from the processing of message orders. When a new message is received, the method {\sf handle()} will be invoked automatically to detect whether the new local context makes the predicate true. Detailed discussions on specific predicate detection algorithms over the STD can be found in \cite{Huang09, Huang12}.

%------------------------------------
\subsubsection{Property Detection over the LAT}

As for property detection over the LAT, the abstract class {\sf AbstractLatticeChecker}, which is a subclass of the {\sf AbstractFIFOChecker}, achieves the maintenance of the LAT. Property detection processes which rely on the LAT (e.g., \cite{Babaoglu96, Yang13, Wei12}) only need to extend the abstract class and implement the specific detection algorithms in the method {\sf check()}, as shown in Fig. \ref{F:AbstractChecker}.

The {\sf AbstractLatticeChecker} constructs the LAT as a concrete subclass of the abstract class {\sf AbstractLattice} in Fig. \ref{F:AbstractChecker}. Different detection algorithms (e.g., \cite{Yang13, Wei12}) may store different historical data on the LAT nodes to achieve incremental detection, as required in Section \ref{sec:Detection of PD-CA}. Thus, we provide the abstract class {\sf AbstractLatticeNode} for different detection algorithms to realize different types of LAT nodes. Furthermore, different LAT construction algorithms (e.g., \cite{Yang13, Yang13a}) are supported by implementing the abstract method {\sf expandLattice()} of the {\sf AbstractLattice}. The construction of the LAT is transparent to the detection algorithms, which shields the detection algorithms from the burden of maintaining the LAT.

When a new local context arrives, the {\sf AbstractLatticeChecker} first updates the LAT incrementally, and then invokes the method {\sf check()} to detect whether the new part of the LAT makes the predicate true.  Detailed discussions on specific LAT maintenance algorithms and predicate detection algorithms over the LAT can be found in \cite{Yang13, Wei12, Yang13a}.

%========================================================================
\section{Simplifying Application Development}\label{sec:develop app}

In this section we discuss how the PD-CA framework simplifies the development of context-aware applications.

%------------------------------------------------------------------------
\subsection{Application Development}\label{sec:application development}
%--
\begin{figure}[tbp]
  \centering
  \includegraphics[width=3.2in]{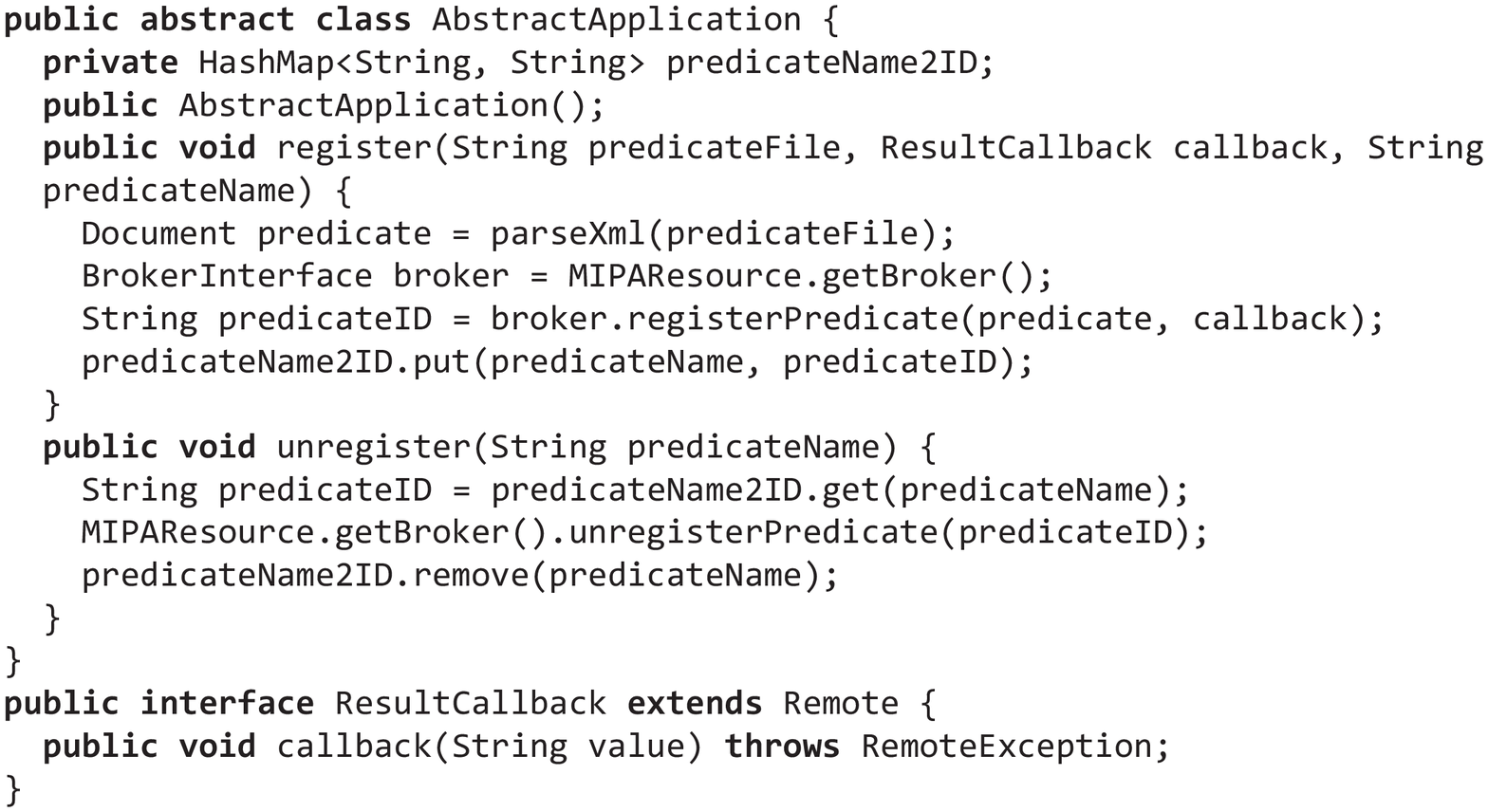}
  \caption{The abstract class AbstractApplication and the interface ResultCallback.}
  \label{F:AbstractApplication}
\end{figure}

Based on MIPA, context-aware applications specify contextual properties to the middleware and conduct the corresponding context-aware behavior when the properties are satisfied. The context-aware adaptation logic of the application is constructed in a condition-action manner. A programming toolkit is provided to facilitate the development of context-aware applications. Specifically, an abstract class {\sf AbstractApplication} is adopted to encapsulate the interactions between the applications and the middleware, and a callback interface {\sf ResultCallback} is introduced for the middleware to notify the applications to conduct context-aware behavior, as shown in Fig. \ref{F:AbstractApplication}. Concrete applications only have to extend the class {\sf AbstractApplication}, specify contextual properties in XML as discussed in Section \ref{sec:Management of the specification}, and implement the corresponding context-aware behavior in the method {\sf callback()} of the {\sf ResultCallback}.

Applications register/unregister contextual properties and the corresponding callbacks to the middleware through the method {\sf register()}/{\sf unregister()}. When a contextual property is detected true, the middleware will trigger the corresponding callback of the application, which conducts the context-aware behavior.

%------------------------------------------------------------------------
\subsection{Exemplar Application}\label{sec:Exemplar application}
%--
\begin{figure}[tbp]
  \centering
  \includegraphics[width=3.2in]{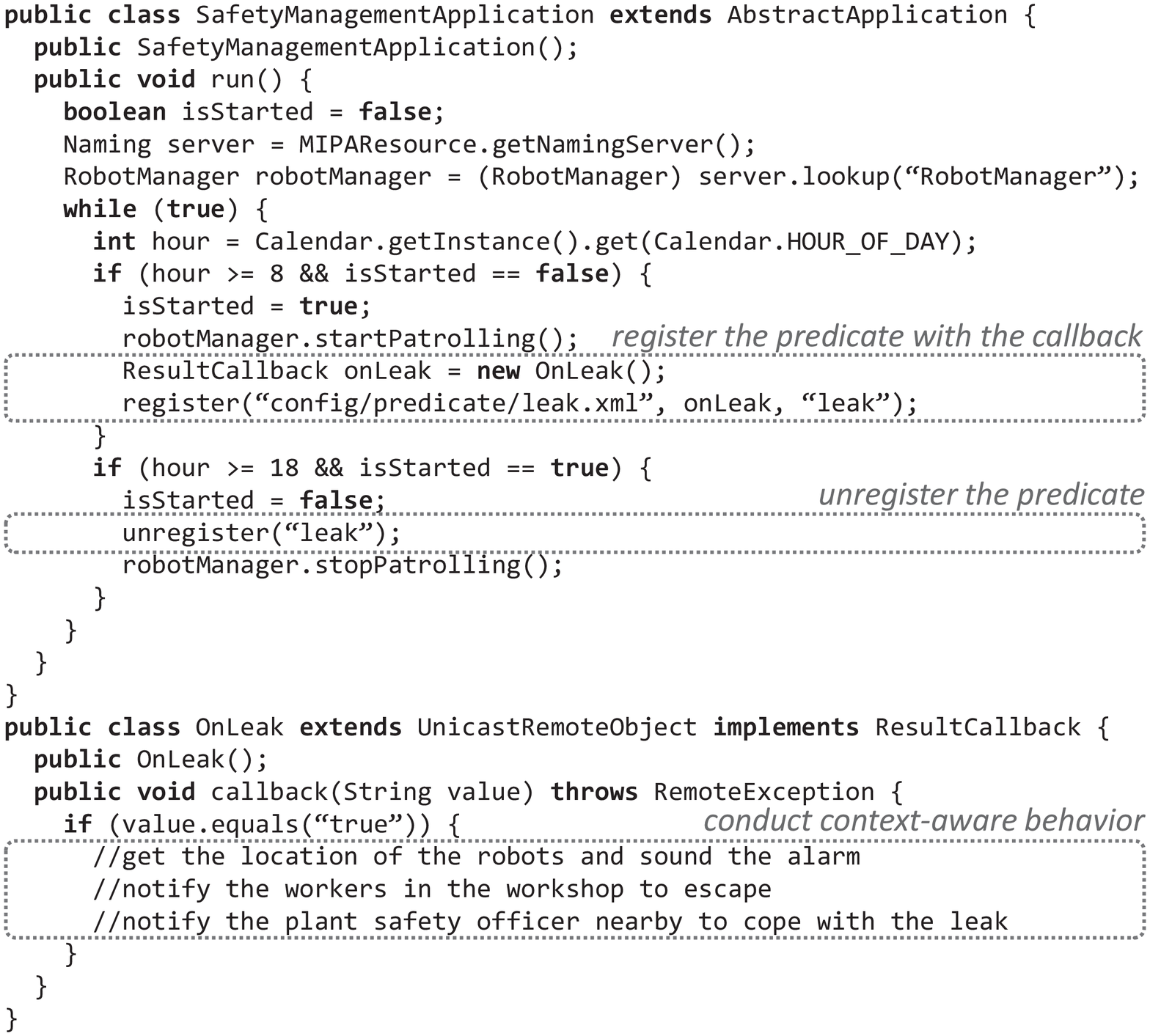}
  \caption{Code for the safety management application.}
  \label{F:Application}
\end{figure}

The best demonstration of the middleware's ability to ease context-aware application development is by examples. We exemplify the development of context-aware applications using the chemical plant safety management application in Section \ref{sec:Case study}.

Assume that the robots patrol the plant from 8:00 to 18:00. At 8:00, the application starts the robots and registers the contextual properties (e.g., $\phi_1$ in Section \ref{sec:Case study}) and the corresponding callbacks to the middleware. When the contextual properties are detected true, the corresponding context-aware behavior will be conducted. At 18:00, the application unregisters the contextual properties and stops the robots. The code of the application is shown in Fig. \ref{F:Application}.

We take the contextual property $\phi_1$ (i.e., all the robots detect hazardous material leak in a workshop) in Section \ref{sec:Case study} as an example. When $\phi_1$ is detected, the safety management application should sound the alarm, notify the workers in the workshop to escape, and notify the plant safety officer nearby to cope with the leak. Property $\phi_1$ in XML is in Fig. \ref{F:specification example} and the corresponding context-aware behavior is implemented in the class {\sf OnLeak} in Fig. \ref{F:Application}. As shown in Fig. \ref{F:Application}, the application registers the property ({\sf config/predicate/onLeak.xml}) and the callback ({\sf onLeak}) to the middleware. When the property is detected true, the method {\sf callback()} of the callback {\sf onLeak} will be invoked automatically to notify the workers and the plant safety officer.

The code of the application is reasonably clean and concise. Under the guidance of our PD-CA framework, MIPA allows the application developers to focus on the development of context-aware behavior, while not having to explicitly cope with the distributed and asynchronous contexts.

%========================================================================
\section{Experimental Evaluations}\label{sec:Experiments}

In this section we evaluate the performance of MIPA. We vary the types of predicates, the number of predicates, and the number of context collecting devices to study the scalability of MIPA. The detection of snapshot predicates and sequence predicates with respect to different degrees of asynchrony in the computing environment has been studied in our previous work \cite{Huang12, Yang13, Wei12}. Please refer to our previous work for detailed evaluations of specific detection algorithms.

%------------------------------------------------------------------------
\subsection{Experiment Design}

We implement MIPA with Java SE 1.6 and run MIPA over JVM 1.6 on a PC with Windows 7 (x64), an Intel Core i5-2400 Quad-Core Processor (3.10GHz), and 8GB RAM. We simulate the plant safety management scenario discussed in Section \ref{sec:Case study} and \ref{sec:Exemplar application}.

We conduct the experiments with two representative predicates: conjunctive predicate $\phi_1$ and regular expression predicate $\phi_4$ in Section \ref{sec:Case study}. We tune the number of predicates and the number of robots, to study the memory consumptions of the property detection server and the context collecting clients, and the response latency of the property detection server. We use $M_{s}$ ($M_{c}$) to denote the average of the memory consumption of the property detection server (all the context collecting clients on the robots) during the lifetime of the experiment, and use $T_{s}$ to denote the average time from the instant when the property detection process is triggered to the instant when the detection finishes.

The application registers a predicate to MIPA every 1 s. The robots sense context data every 400 ms. We simulate the context data of the robots with the Poisson distribution. The average time of the local activities (where the local predicate is true) on the robots is 5 mins, and the average interval between the activities (where the local predicate is false) is 1 min. The context data of each robot is up to 16,000. We model the communication delay between the robots by exponential distribution with average delay 10 ms. The lifetime of the experiment is up to 2 hours.

%------------------------------------------------------------------------
\subsection{Evaluation Results}

%------------------------------------
\subsubsection{Performance with Conjunctive Predicates}

We first evaluate the performance of MIPA with conjunctive predicate $\phi_1$ in Section \ref{sec:Case study}. We tune the number of conjunctive predicates from 1 to 1,000 and the number of robots from 10 to 40.

\begin{figure}[tbp]
  \centering
  \includegraphics[width=3.2in]{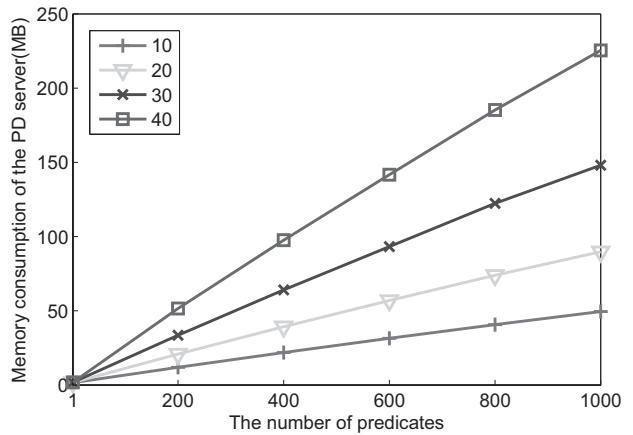}
  \caption{The average memory consumption of the property detection server under conjunctive predicates.}
  \label{F:wcp server space}
\end{figure}

As shown in Fig. \ref{F:wcp server space}, when fixing the number of robots, $M_{s}$ increases linearly with the number of predicates. When fixing the number of predicates, $M_{s}$ increases slowly with the number of robots. The slow increase of $M_{s}$ is because the detection algorithm for conjunctive predicates is space-efficient \cite{Huang12}. When the number of robots is 10, 20, 30, and 40, $M_{s}$ of 1,000 predicates is 49.5 MB, 89.8 MB, 148.1 MB, and 225.4 MB, respectively. Thus, $M_{s}$ is very small when detecting conjunctive predicates.
%--
%--
\begin{figure}[tbp]
  \centering
  \includegraphics[width=3.2in]{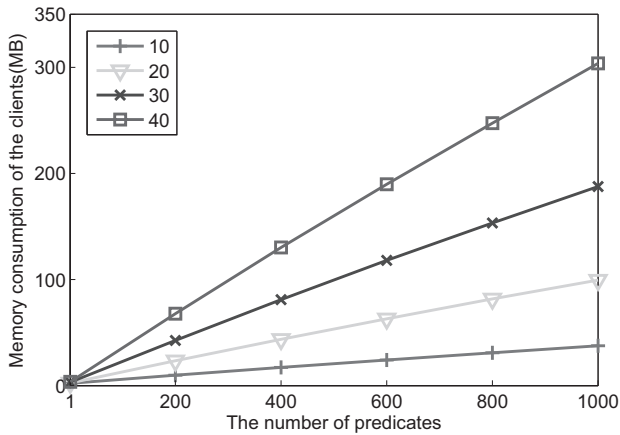}
  \caption{The average memory consumption of the context collecting clients under conjunctive predicates.}
  \label{F:wcp client space}
\end{figure}

As shown in Fig. \ref{F:wcp client space}, when fixing the number of robots, $M_{c}$ increases linearly with the number of predicates. When fixing the number of predicates, $M_{c}$ increases slowly with the number of robots. When the number of robots is 10, 20, 30, and 40, $M_{c}$ of 1,000 predicates (in total 10,000, 20,000, 30,000, and 40,000 context collecting processes) is 37.4 MB, 99.4 MB, 187.7 MB, and 303.7 MB, respectively. That is, for each robot, the average space cost is 3.7 MB, 5.0 MB, 6.3 MB, and 7.6 MB, respectively. Thus, $M_{c}$ is acceptable on resource-constrained context collecting devices, such as mobile robots and phones.
%--
\begin{figure}[tbp]
  \centering
  \includegraphics[width=3.2in]{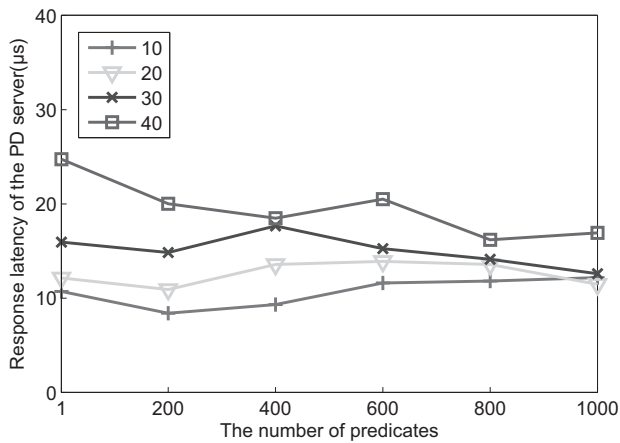}
  \caption{The average response latency of the property detection server under conjunctive predicates.}
  \label{F:wcp latency}
\end{figure}

As shown in Fig. \ref{F:wcp latency}, when fixing the number of robots, $T_{s}$ remains almost the same. When fixing the number of predicates, $T_{s}$ increases slowly with the number of robots. When we tune the number of predicates from 1 to 1,000, $T_{s}$ remains pretty small (within 30 $\mu$s). The reason is that the detection algorithm for conjunctive predicates is time-efficient \cite{Huang12}.

%------------------------------------
\subsubsection{Performance with Regular Expression Predicates}

We then evaluate the performance of MIPA with regular expression predicate $\phi_4$ in Section \ref{sec:Case study}. We tune the number of regular expression predicates from 1 to 100 and the number of robots from 3 to 6.
%--
\begin{figure}[tbp]
  \centering
  \includegraphics[width=3.2in]{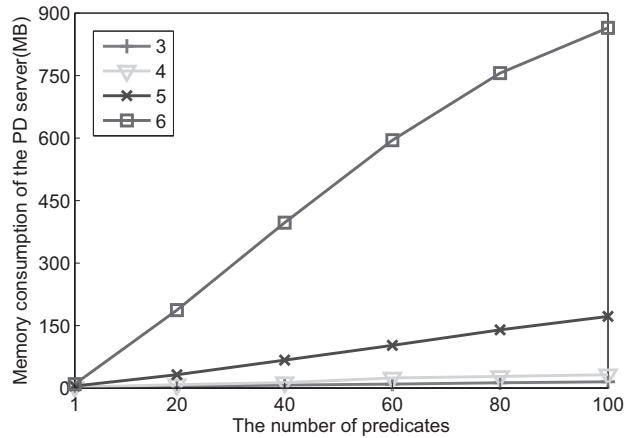}
  \caption{The average memory consumption of the property detection server under regular expression predicates.}
  \label{F:re server space}
\end{figure}

As shown in Fig. \ref{F:re server space}, when fixing the number of robots, $M_{s}$ increases almost linearly with the number of predicates. When fixing the number of predicates, $M_{s}$ increases exponentially with the number of robots. This is in accordance with the exponential detection algorithm for regular expression predicates \cite{Yang13}. When the number of robots is 3, 4, 5, and 6, $M_{s}$ of 100 predicates is 15 MB, 32 MB, 172 MB, and 864 MB, respectively. $M_{s}$ of regular expression predicates is much more than that of conjunctive predicates shown in Fig. \ref{F:wcp server space}. Thus, the detection of regular expression predicates can only work well in a small scale of context collecting devices.
%--
\begin{figure}[tbp]
  \centering
  \includegraphics[width=3.2in]{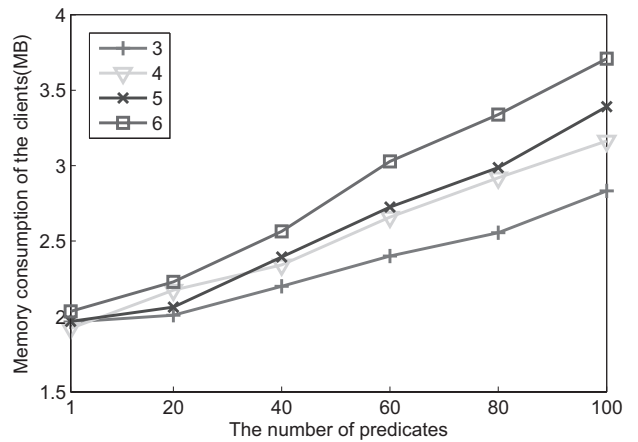}
  \caption{The average memory consumption of the context collecting clients under regular expression predicates.}
  \label{F:re client space}
\end{figure}

As shown in Fig. \ref{F:re client space}, when we tune the number of predicates from 1 to 100, $M_{c}$ remains within the range of 1.5 MB to 4 MB. When fixing the number of robots, $M_{c}$ increases linearly and slowly with the number of predicates. When the number of robots is 3, 4, 5, and 6, $M_{c}$ of 100 predicates is 2.8 MB, 3.2 MB, 3.4 MB, and 3.7 MB, respectively. $M_{c}$ increases almost linearly with the number of robots, and costs much less than $M_{s}$. This is because, although the detection algorithm for regular expression predicates on the property detection process is exponential (as shown in Fig. \ref{F:re server space}), the algorithm on the context collecting process is linear \cite{Yang13}. $M_{c}$ is acceptable on resource-constrained context collecting devices.
%--
\begin{figure}[tbp]
  \centering
  \includegraphics[width=3.2in]{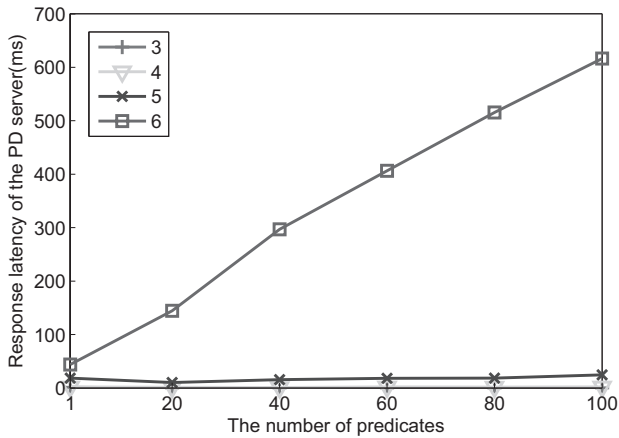}
  \caption{The average response latency of the property detection server under regular expression predicates.}
  \label{F:re latency}
\end{figure}

As shown in Fig. \ref{F:re latency}, when fixing the number of robots, $T_{s}$ increases linearly with the number of predicates. When the number of robots is 3, 4, and 5, $T_{s}$ of 100 predicates is 0.2 ms, 1.9 ms, 24.5 ms, respectively. When the number of robots is 6, $T_{s}$ of 100 predicates increases quickly to 616.4 ms. $T_{s}$ increases exponentially with the number of robots. This is in accordance with the exponential detection algorithm for regular expression predicates \cite{Yang13}. Thus, the detection of regular expression predicates can only work well in a small scale of context collecting devices.

Although the performance with respect to regular expression predicates is not as scalable as that with respect to conjunctive predicates, in pervasive computing scenarios, the number of context collecting devices is usually on a small scale \cite{Yang13}. Thus the performance of detection of regular expression predicates is often acceptable.

%%------------------------------------------------------------------------
%\subsection{Lessons Learned}\label{}
%
%\YH{what to say here? discuss the logical flow and the outline first.}
%
%Our predicate detection-based context-aware approach focuses on event ordering on purely asynchronous systems and the expressiveness is limited by the asynchronous model. In general context-aware scenarios, our approach can be combined with other context-aware approach to better support context-awareness. For example, our approach and approaches working on synchronous models can be combined to work in approximately synchronous scenarios. Our approach focuses on detection of contextual properties concerning on event ordering and approaches which work on synchronous models focus on detection of timed contextual properties.
%
%The predicate detection theory has an assumption on available fault-tolerance mechanisms, which is not supported by our middleware yet. To work in realistic scenarios, dedicated fault-tolerance modules should be developed to guarantee the assumption.

%========================================================================
\section{Related Work}\label{sec:Related work}

Context-aware computing has been extensively studied in the literature. From the point of view of this work, we mainly discuss two types of related work: the key enabling techniques and the software engineering methodologies for context-aware computing.

As for the key enabling techniques, tuple-space-based approaches provide an egocentric view of the dynamic environment for an individual application. Contexts are abstracted as data items stored in a distributed tuple space. The application achieves context-awareness by interacting with the contexts in the tuple space \cite{Murphy06, Julien06, Mamei09}. LIME enables mobile coordination by abstracting communication into the tuple space \cite{Murphy06}. EgoSpaces enables each agent to specify its egocentric view of the network \cite{Julien06}. TOTA, augmenting tuple spaces with reactive capabilities, provides a push-based interaction mechanism, i.e., tuples are propagated from a reference node based on context properties in a manner similar to content-based multicast \cite{Mamei09}. Our approach, instead of maintaining available contexts for the application, enables the application to declare its concern using hight-level predicates. The middleware maintains contexts related to the predicates, conducts predicate detection, and notifies the application of the satisfaction of the predicates.

Query-based approaches provide a database-like abstraction of the computing environment of an application. The collection of contextual information available to a particular application is abstracted as a global virtual data repository that reflects the continuously changing state of the application's environment. The application achieves context-awareness by querying the ``database''. PerLa \cite{Schreiber12} provides a SQL-like language which is able to manage heterogeneous devices and support most existing sampling modes. Payton et al. \cite{Payton10} develop a self-assessing query processing protocol for dynamic environments, which not only delivers a query result but also labels that result with the achieved consistency semantics. Our approach also provides a specification for an application to express contextual properties of its interest. Though the specification of predicates in our PD-CA framework has less expressiveness compared to database query languages, our work mainly targets at coping with the asynchrony of the computing environment.

Pub/sub-based approaches, such as Solar \cite{Chen08} and STEAM \cite{Meier10}, enable a number of subscribers to continuously retrieve events from a number of publishers. These approaches develop efficient multicast-based routing and in-network event filtering techniques. Our predicate-detection-based approach is analogous to pub/sub-based approaches in that the application subscribes the predicates to the middleware, while the middleware subscribes the consisting local predicates to the distributed context collecting devices. Our work is a special case of the general pub/sub approach, but it is dedicated to coping with the asynchrony of the computing environment.

Recently, many software methodologies for the development of context-aware applications are proposed. Anind K. Dey \cite{Dey00} identifies a design process for building context-aware applications and provides two programming abstractions to facilitate the design of such applications. The Context Toolkit is built to support the design process and the programming abstractions. Henricksen et al. \cite{Henricksen06} introduce a graphical context modelling approach, a preference model for representing context-dependent requirements, and two programming models for developing context-aware applications. A generic software engineering process is also proposed when building context-aware applications using these tools. Cassou et al. \cite{Cassou12} introduce a design methodology as well as corresponding tool support for pervasive applications. Testing and maintenance are also discussed.

We propose the PD-CA framework to enable context-awareness in asynchronous pervasive computing environments enriched with C$^3$S devices. The PD-CA framework employs the predicate detection theory to address the challenge of the asynchrony. PD-CA also includes a design process for developing applications aware of asynchronous environments.

%========================================================================
\section{Conclusion and Future Work}\label{sec:Conclusion}

In this paper, we study how to enable context-awareness in asynchronous pervasive computing environments enriched with C$^3$S devices. We propose the PD-CA framework, which utilizes the predicate detection theory to cope with the intrinsic asynchrony of the pervasive computing environment. PD-CA also proposes a design process for developing context-aware applications. Under the guidance of PD-CA, we design and implement the MIPA middleware. We also demonstrate how MIPA simplifies the development of context-aware applications.

This work is being expanded in various directions. To better work in real scenarios, fault-tolerance in the PD-CA framework must be strengthened. The most urgent issues to be addressed include the crash failure of context collecting devices and the link failure of wireless communications. The PD-CA framework also needs to support multiple time models in the future. For example, the partially synchronous model should be integrated in the PD-CA framework to cover more pervasive computing scenarios.

%========================================================================
\section*{Acknowledgements}

This work is supported by the National Natural Science Foundation of China (No. 61272047, 91318301).

%--
\bibliographystyle{IEEEtran}
\bibliography{IEEEabrv,MIPA}

\end{document}